\documentclass{emulateapj}
\usepackage{graphicx}
\begin{document}
\title{Assessing Pulsar Timing Array Sensitivity to\\ Gravitational Wave Bursts With Memory}
\author{
	D. R. Madison, J. M. Cordes, S. Chatterjee\\
	Department of Astronomy and Center for Radiophysics and Space Research,\\ Cornell University, Ithaca, NY 14853, USA	
	}
\begin{abstract}
Highly energetic astrophysical phenomena like supermassive black hole binary (SMBHB) mergers are predicted to emit prodigious amounts of gravitational waves (GWs).  An anticipated component of the gravitational waveform known as ``memory" is permanent and non-oscillatory.  For SMBHB mergers, the memory is created primarily during the most violent moments of the inspiral immediately preceding the final plunge and ring-down when the strongest gravitational fields are at work and the non-linearities of general relativity are most pronounced.  The essentially time-domain nature of memory makes it forbiddingly difficult to detect with ground based GW detectors, leaving pulsar timing array (PTA) experiments as the most promising means by which it may be detected and studied.  In this paper, we discuss how GW bursts with memory (BWMs) influence pulsar timing experiments and develop methods to assess how sensitive modern timing efforts are to such GW events.  We discuss how PTA searches for BWMs can be used to constrain the rate of BWMs and how these constraints relate to information regarding the population of SMBHBs. 
\end{abstract}
\maketitle

\section{Introduction}
Bursts of gravitational waves (GWs) are predicted to contain non-oscillatory components that lead to permanent deformations of space-time.  These deformations are created when gravitons escape gravitationally radiating systems and cause their mass-energy quadrupole moments to have permanently non-vanishing second time derivatives.  Because the non-oscillatory components have magnitudes that are sensitive to the entire history of the source, they are commonly referred to as ``memory'' \citep{s77,bp79,bt87,c91,bd92,f09}.  

Precision pulsar timing has long been recognized as a means by which extremely low-frequency GWs might be detected \citep{s78,d79} and the potential of combining the timing data from an array of pulsars (a pulsar timing array or PTA) for enhanced GW sensitivity and source characterization has been appreciated for nearly as long \citep{hd83,fb90}.  The European Pulsar Timing Array \citep[EPTA;][]{kc13}, the Parkes Pulsar Timing Array \citep[PPTA;][]{h13}, and the North American Nanohertz Observatory for Gravitational Waves \citep[NANOGrav;][]{m13} are collaborations of scientists working to realize the PTA concept, detect GWs, and characterize their sources.  These PTAs are being used to search for a stochastic background (SB) of GWs from an ensemble of supermassive black hole binaries (SMBHBs) spread throughout the universe \citep{vlj+11,dfg+13,src+13}, coherent GWs from individual SMBHBs \citep{yhj+10,lwk+11,esc12}, and generic GW bursts \citep{fl10}.  Recently, a great deal of consideration has gone into the detection of GW bursts with memory (BWMs) by PTAs \citep{s09,pbp10,vl10,cj12}.  

All PTA efforts use the same fundamental strategy.  Each pulsar in an array of particularly rotationally stable millisecond pulsars (MSPs) is observed on a regular basis to measure high-precision pulse times of arrival (TOAs).  Pulsars emit pulses at nearly uniform intervals that slowly grow as the pulsars shed angular momentum.   The regularity with which these pulses arrive at Earth is modulated by effects such as binary motion of the pulsar, the motion of the Earth about the Sun, and fluctuations in the amount of ionized interstellar plasma between the Earth and the pulsar.  The TOA variations associated with these processes can be analytically described and precisely modeled.  With these techniques, several MSPs produce timing residuals (the deviations of measured TOAs from the predictions of a timing model) over 5 to 10 year spans that are noise-like with an rms of just a few tens of nanoseconds.  Dozens more MSPs produce timing residuals with an rms of a few hundred nanoseconds over similar spans.  With the TOAs of some pulsars deviating so slightly from the predictions of models, small effects that have not been incorporated into the models may soon become measurable.  Examples are improper modeling of the solar system \citep{chm+11}, errors in terrestrial time standards \citep{hcm+12}, or, as is the concern of this paper, GW memory, a telltale indicator of SMBHB mergers.

In Section 2, we describe the signature of a BWM in PTA data.  In Section 3, we discuss techniques for assessing PTA sensitivity to BWMs and searching for them.  In Section 4, we describe how PTA searches for BWMs, in the event of a non-detection, can be used to constrain the rate of BWMs and how these constraints can inform us about the population of SMBHBs.  Finally, in Section 5 we summarize and discuss our work and discuss prospects for the future of BWM investigations by PTAs.

%%%%%%%%%%%%%%%%%%%%%%%%%%%%%%%%%%%%%%%%%%%%%%%%%%%%%%%%%%%%%%%%%%%%%%%%%%%%%%%%%%%%%%%%%%%%%%%%%

\section{GW Bursts With Memory}

\begin{figure}[tbp]
\begin{center}
\includegraphics[height = 85mm, width = 70mm]{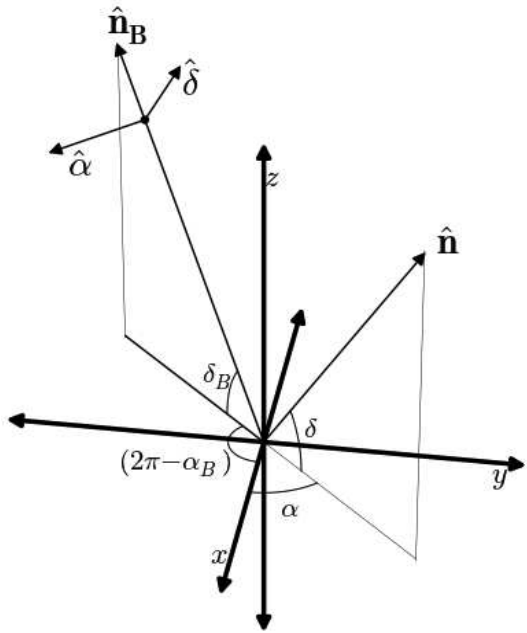}
\caption{A schematic diagram depicting several vectors relevant for determining the trigonometric weight that influences the magnitude of the influence of a BWM in a pulsar's timing residuals. The observer is at the origin. The vector ${\bf \hat{n}}$ points towards the pulsar while ${\bf \hat{n}_B}$ points toward the GW source.  The vectors ${\bf \hat{\alpha}}$ and ${\bf \hat{\delta}}$ span the plane normal to the propagation direction of the wave. }
\end{center}
\end{figure}

The sources of BWMs most likely to be detectable by PTAs are the mergers of SMBHBs with mass ratios near unity.  For merging black holes of mass $M_1$ and $M_2$ ($M_1\geq M_2$), the dimensionless strain of the memory has an amplitude
\begin{eqnarray}
h_+^{\rm (mem)}\approx\frac{(1-\sqrt{8}/3)}{24}\frac{G\mu}{c^2D}\sin^2{\cal I}(17+\cos^2{\cal I})\nonumber\\\times\left[1+{\cal O}(\mu^2/M^2)\right],
\end{eqnarray}
where $M=M_1+M_2$, $\mu=M_1M_2/M$ is the reduced mass, ${\cal I}$ is the inclination angle just prior to merger, and $D$ is the comoving distance to the source \citep{f09,lcz+10,pr11}.  The ``$\times$'' polarization of the memory vanishes for circularized binaries.  The higher order contributions to Equation (1) depend on the orientation of the angular momenta of the BHs just prior to the final plunge and may not be small.  Nonetheless, if we neglect the higher order contributions to Equation (1), an equal-mass merger of two $10^9 M_{\odot}$ BHs a distance of 1 Gpc from the Earth is predicted to produce memory with an amplitude $h_+^{\rm (mem)}\approx 10^{-15}$ with optimal beaming.

It is during the final stages of the BH merger when several percent of the system's total rest-mass energy is radiated as GWs that the memory grows most precipitously.  The timescale $\tau$ for the memory to undergo this final growth is roughly equal to the time light takes to circumnavigate the event horizon of the post-merger BH: $\tau\approx 2\pi R_S/c$ where $R_S$ is the Schwarzschild radius \citep{cj12}.  If the merger product is a $10^9~M_\odot$ BH, $\tau$ is approximately one day.  For mergers occurring at redshift $z$, the observed rise-time will increase by $(1+z)$ \citep{vl10}.  

\subsection{Influence of BWMs in Pulsar Timing Data}

Consider a linearly polarized GW encountering the Earth from a source in the direction ${\bf \hat{n}_B}$ at right ascension $\alpha_B$ and declination $\delta_B$.  Call the propagation direction of the wave ${\bf \hat{k}}$ and the principal polarization direction of the wave ${\bf \hat{\psi}}={\bf \hat{\delta}}\cos{\psi}+{\bf \hat{\alpha}}\sin{\psi}$ where ${\bf \hat{\delta}}$ and ${\bf \hat{\alpha}}$ form an orthonormal basis spanning the plane normal to ${\bf \hat{k}}$ and $\psi$ is the polarization angle of the wave increasing from ${\bf \hat{\delta}}$ towards ${\bf \hat{\alpha}}$.  For a pulsar a distance $l$ from Earth located along the direction indicated by ${\bf \hat{n}}$ (many of the vectors we have discussed are depicted in Figure 1) the GW influences the apparent pulsation frequency, $\nu$, of a pulsar as follows \citep{ew75,w87,h91}:
\begin{eqnarray}
\frac{\Delta\nu(t)}{\nu}&=&B(\theta,\phi)\left[h(t)|_E-h(t-t_l)|_{\bf n}\right],\\
B(\theta,\phi)&=&\frac{1}{2}\cos{(2\phi)}(1-\cos{\theta}),\\
t_l&=&\frac{l}{c}\left(1+\cos{\theta}\right),\\
\theta&=&\cos^{-1}{({\bf \hat{n}\cdot\hat{k}})},\\
\phi&=&\tan^{-1}{\left(\frac{{\bf \hat{n}\cdot\hat{\alpha}}}{{\bf \hat{n}\cdot\hat{\delta}}}\right)}-\psi.
\end{eqnarray}
In words, $\theta$ is the angle between the line of sight to the pulsar and the propagation direction of the wave and $\phi$ is the angle between the principal polarization vector of the wave and the projection of the pulsar line of sight onto the plane normal to the wave propagation direction.  The terms in Equation (2) are to be evaluated at the location of the Earth ($E$) and the pulsar (${\bf n}$) as indicated.  For fixed ${\bf \hat{k}}$ and ${\bf \hat{n}}$, if one averages over polarization angle, it can be shown that
\begin{eqnarray}
\langle B^2(\theta,\phi)\rangle^{1/2}_\psi = \frac{1}{\sqrt{8}}(1-\cos{\theta}).
\end{eqnarray}
For fixed ${\bf \hat{k}}$ and $\psi$, if one averages ${\bf \hat{n}}$ over the entire sky, it can be shown that 
\begin{eqnarray}
\langle B^2(\theta,\phi)\rangle^{1/2}_\Omega = \frac{1}{\sqrt{6}}.
\end{eqnarray}

Since the typical measurement cadence for TOAs is approximately a month and the anticipated rise time for the memory component of a GW signal is at most a few days, we will ignore the detailed shape of the growing memory signal and model it approximately as a simple step function of amplitude $h_B$ that turns on at a time $t_0$.  As pulsar timing experiments measure the rotational phase of pulsars, it is the integral of the fractional frequency change that determines the perturbation from the BWM signal on the observed TOAs.  Thus, the timing perturbation associated with a BWM is
\begin{eqnarray}
\Delta t(t) = h_BB(\theta,\phi)[&(t-t_0)\Theta(t-t_0)-&(t-t_1)\Theta(t-t_1)],\nonumber\\
\end{eqnarray}   
where $t_1=t_0+t_l$.  This is the same BWM signal model used by \citet{vl10}, \citet{pbp10}, and \citet{cj12}.  One additional assumption that is important in the derivation of this expression is that $l\ll D$.  When this assumption is valid, the amplitude of the memory does not change appreciably as it propagates between the Earth and the pulsar and differences in $\theta$ and $\phi$ at the Earth and the pulsar associated with a burst source at a finite distance are negligibly small.  

The first term in the brackets of Equation (9) is referred to as the Earth term and the second is referred to as the pulsar term.  The time $t_0$ is when the memory wavefront strikes the Earth.  The necessarily later time $t_1$ is when the memory wavefront is observed to strike the pulsar.  Since $l$ will typically be on the order of a kiloparsec, unless the pulsar and the BWM source have very little angular separation ($\theta\approx\pi$), $t_1$ will lag $t_0$ by hundreds to thousands of years.  Because of this, if a BWM is detected in a timing data set, which at the very most has a length of a couple of decades, it will likely only be detected in the Earth term or the pulsar term, but not both.

The timing perturbation associated with a BWM is thus a linearly growing ramp function that begins to grow when either the Earth or pulsar term is activated.  If a BWM occurs during a timing campaign, this functional form will not appear in the timing residuals after the timing model is fit to the meaured TOAs.  It is common practice to iteratively refine the parameters of a timing model as new data is acquired to minimize the rms of the residuals.  Consider an initial timing model for some pulsar that yields a phase-connected timing solution and is parameterized by a vector ${\bf p}$ of $m$ parameters (see \citet{lk05} for more on the basics of pulsar timing).  It is assumed that the $n$ timing residuals with this initial timing model (${\bf R_{\rm pre}}$) are non-zero because of additive noise (${\bf \Sigma}$) and small deviations in the timing model parameters away from their maximum likelihood values (${\bf \delta p}$; small enough deviations that the residuals fluctuate linearly in response to changes in the timing model parameters), i.e. ${\bf R_{\rm pre}} = {\bf M\delta p}+{\bf \Sigma}$ where ${\bf M}$ is the $n\times m$ design matrix describing how the residuals change with changes in the timing model parameters.  If ${\bf C} = \langle{\bf\Sigma\Sigma^T}\rangle$ is the covariance matrix of the noise, and if the noise is Gaussian, the maximum likelihood estimate for the optimal modifications to the timing model is \citep{g10}
\begin{eqnarray}
{\bf\delta{\hat p}} = ({\bf M}^T{\bf C}^{-1}{\bf M})^{-1}{\bf M}^T{\bf C}^{-1}{\bf R_{\rm pre}},
\end{eqnarray}
and the timing model parameter covariance matrix is
\begin{eqnarray}
{\bf C_p}=({\bf M}^T{\bf C}^{-1}{\bf M})^{-1}.
\end{eqnarray}
Such pulsar timing techniques are used by all existing PTAs to search for GW signatures \citep{vlj+11,dfg+13,src+13}.

The ramp function described in Equation (9) will be covariant with some parameters of any timing model, mainly the pulsation frequency $\nu$ and its derivative $\dot{\nu}$.  If the signature of a BWM is present in a set of timing residuals, fitting for $\nu$ and $\dot{\nu}$ will subtract from it a quadratic that optimally matches it in a least-squares sense.  A BWM signal has lesser covariances with other fit parameters, so the signature of a BWM in post-fit residuals intimately depends on the details of the timing model.  We will explore this in detail in Section 3.

\subsection{BWMs vs. Glitches}

Many pulsars are known to undergo spontaneous sudden pulsation frequency changes known as glitches.  Glitches are attributed to complex behavior in the interior of the neutron star or in its magnetosphere and tend to display additional phenomenology that can, in principle, differentiate them from BWMs.  Most glitches in $\nu$ have associated glitches in $\dot{\nu}$, and the timing parameters altered by the glitch often fully or partially relax back to their pre-glitch values on an exponential timescale of months \citep{ymh+13}.  None of this behavior will accompany a BWM.  Additionally, while a BWM can lead to an increase or decrease in $\nu$, until very recently, all observed glitches have caused an increase in $\nu$.  One so-called ``anti-glitch" has been observed in the magnetar 1E 2259+586 \citep{akn+13}.  

Most of the pulsars that are known to glitch are young, high-magnetic-field, long-period canonical pulsars (CPs) that display complicated timing behavior and are not candidates for precision timing experiments.  Only some MSPs are candidates for high precision timing.  However, even MSPs have rarely demonstrated glitches.  \citet{cb04} observed a glitch in B1821-14.  This pulsar is among the youngest known MSPs and displays significant amounts of red timing noise.  The B1821-14 glitch was accompanied by a marginally significant glitch in $\dot{\nu}$.  The fractional frequency change associated with this glitch was $10^{-11}$, two orders of magnitude smaller than the smallest glitch that had been observed in CPs to that date.  If this glitch were attributed to a BWM (despite the observed glitch in $\dot{\nu}$), the source would have to be tremendous--something like a near-equal-mass merger of a $10^{10}$ $M_\odot$ edge-on SMBHB just 10~Mpc from the Milky Way.  Even the smallest observed glitches tend to have such large magnitudes that they cannot be readily ascribed to BWM events created by the merger of astrophysical systems.

A BWM detected in the pulsar term of a single pulsar may be difficult to differentiate from a micro-glitch in that pulsar.  However, if an array of pulsars is monitored, a simultaneous detection in the shared Earth term would be distinguishable from any pulsar-specific phenomena.  Moreover, the frequency change in all of the pulsars being monitored would have a quadrupolar amplitude pattern consistent with a GW from a certain direction and with a certain polarization as determined by $B(\theta,\phi)$ and the orientation of the pulsars on the sky.  The residuals of the array can be combined and weighted appropriately so as to enhance the PTA's sensitivity to BWMs and to extract information regarding the source direction and polarization.\\

%%%%%%%%%%%%%%%%%%%%%%%%%%%%%%%%%%%%%%%%%%%%%%%%%%%%%%%%%%%%%%%%%%%%%%%%%%%%%%%%%%%%%%%%%%%%%%%%%

\section{PTA Sensitivity to BWMs}
In this section, we discuss methods for assessing the sensitivity of a PTA to BWMs appearing in either pulsar terms or Earth terms.  We also discuss a method for detecting a BWM in a timing data set for a single pulsar and the prospects for correctly estimating the amplitude and epoch of the burst.  

\subsection{BWM Sensitivity with Idealized PTA Data}

In the case of uniform TOA measurement cadence ($\Delta t$), uniform TOA uncertainty from white Gaussian noise ($\sigma_{\rm TOA}$) and simple timing models in which only $\nu$ and $\dot{\nu}$ are fit for, \citet{vl10} have analytically calculated the minimum amplitude BWM that can be detected in a timing data set of length $T$ (with 1-$\sigma$ certainty):
\begin{eqnarray}
h_{\rm min} &=& 2\sqrt{3}B^{-1}(\theta,\phi)\sigma_{\rm TOA}(\Delta t)^{1/2}T^{-3/2}f(t/T),\\
f(x)&=&\left[x^3(1-x)^3(15x^2-15x+4)\right]^{-1/2}.
\end{eqnarray} 
\citet{vl10} have also considered the situation in which the residuals of many pulsars are coherently combined to enhance the sensitivity to BWMs in the Earth term. They show that in the ideal case of $N_P$ pulsars with positions ${\bf \hat{n}_i}$, simple timing models in which only $\nu$ and $\dot{\nu}$ are fit for, uniform TOA uncertainties $\sigma_i$, and the same observation span and uniform observing cadence, the Earth-term BWM sensitivity becomes
\begin{eqnarray}
h_{\rm min} &=& 2\sqrt{3}\sigma_{\rm eff}(\Delta t)^{-1/2}T^{-3/2}f(t/T),\\
\sigma_{\rm eff}&=&\left(\sum_{i=1}^{N_P}\sigma_i^{-2}B^2(\theta_i,\phi_i)\right)^{-1/2}.
\end{eqnarray}
We take these idealized results as benchmarks to which we will compare several of our results.

%%%%%%%%%%%%%%%%%%%%%%%%%%%%%%%%%%%%%%%%%%%%%%%%%%%%%%%%%%%%%%%%%%%%%%%%%%%%%%%%%%%%%%%%%%%%%%%%%

\subsection{BWM Sensitivity with Realistic Data Sets}

Realistic pulsar timing data fail to satisfy many of the assumptions used to derive Equations (12)-(15).  Though effort is made to measure TOAs at a regular cadence, unforeseen telescope maintenance or competition for telescope time often prevent this.  Interstellar scintillation--intensity modulations of a pulsar's flux density caused by scattering in the interstellar medium (ISM)--can dramatically alter the signal-to-noise ratio (S/N) of a pulsar causing the TOA uncertainty to fluctuate between observations \citep{cs10}.  Timing models can be complex, benefitting from fitting many more parameters than just $\nu$ and $\dot{\nu}$.  The noise contaminating timing residuals is often not entirely white as many pulsars display some amount of red intrinsic spin noise \citep{sc10}.  We aim to assess how much these features of timing data influence BWM sensitivity.

\subsubsection{Data Model}
We have taken the NANOGrav data described in \citet{dfg+13} as a realistic model for our simulated PTA.  Members of NANOGrav observed 17 MSPs on a roughly monthly basis over an approximately 5 year span with the Green Bank Telescope (GBT) and Arecibo.  One pulsar, J1713+0747, was observed with both telescopes, but otherwise, pulsars visible to Arecibo were observed with Arecibo because of its greater sensitivity and those not visible to Arecibo were observed with the fully-steerable GBT.  Identical backends at the two telescopes, the Astronomical Signal Processor (ASP) and the Green Bank Astronomical Signal Processor (GASP), were used to conduct all observations. These backends carried out real-time coherent dedispersion over a 128 MHz band (divided into 32 4 MHz channels).  

Many TOAs were reported from each observation owing to the channelization of the wide band.  Because of stable but unaccounted for pulse profile evolution across the band, constant offsets between TOAs from different channels were fit for.   At Arecibo, pulsars were observed at two widely-separated frequencies (typically 820 MHz and 1.4 GHz) within one day; at the GBT, pulsars were observed at a second frequency within one week of the first observation.  With this multi-frequency data, TOA fluctuations caused by epoch-to-epoch dispersion measure (DM) variation were fit for.  In addition to parameters describing frequency-dependent TOA variations, $\nu$ and $\dot{\nu}$, up to 5 astrometric parameters, and any important binary parameters (some relativistic) were fit for.

We have excluded several pulsars in the \citet{dfg+13} data set from our analysis for the following reasons.  Three pulsars in the data set (J1853+1308, J1910+1256, and B1953+29), have insufficient multi-frequency data to correct for DM fluctuations. This introduces low-frequency correlated noise to the residuals that cannot be modeled appropriately without extensive multi-frequency follow-up observations.  Two of the pulsars (J1600$-$3053 and B1953+29) have shorter data spans than the others by approximately a factor of two.  Shorter timing spans do not in principle disqualify these pulsars from a search for BWMs, but with such small timing spans, the sensitivity to BWMs is automatically reduced by a factor approximately equal to $2^{-3/2}$ (see Equation 12) compared to what the sensitivity would be with a full five years of timing data with the residual rms and observing cadence held fixed. These pulsar data sets will thus not likely prove to be very sensitive probes of BWMs when compared to other pulsars with five full years of timing data. Additionally, low-frequency red noise with a power spectrum $P(f)\propto f^{-\gamma}$ in the timing residuals of these pulsars with approximately two-year observing spans will be highly covariant with the timing models, especially for $2<\gamma<5$.  This covariance is greatly reduced once five full years of timing data are available \citep{mcc13}. These spectral types of red noise are very descriptive of noise processes seen in many pulsars and are expected to be present in MSP timing residuals at some amplitude \citep{sc10}. 

From the selected NANOGrav data sets, we have generated and analyzed simulated sets of TOAs that have the same observation epochs, TOA uncertainties, and underlying timing models as the real data, but consist of only noise rather than the actual NANOGrav residuals.  We have done this to control the character of the noise in the residuals and assure they contain no unaccounted for deterministic signals or systematic effects.  We can also create many realizations of these simulated TOAs to study their ensemble behavior.  We have used the TEMPO2 pulsar timing package to create and analyze these simulations \citep{hem06}.

%%%%%%%%%%%%%%%%%%%%%%%%%%%%%%%%%%%%%%%%%%%%%%%%%%%%%%%%%%%%%%%%%%%%%%%%%%%%%%%%%%%%%%%%%%%%%%%%%

\subsubsection{Assessing Pulsar Term Sensitivity with Design\\ Matrices and Simulated Data}
To assess how sensitive realistic pulsar timing data sets are to BWMs, with their variable TOA uncertainties, non-uniform observing cadence, and complicated timing models, we have utilized Equation (11).  We have primarily considered white noise models.  The assumption of white noise will yield the most optimistic sensitivity estimates consistent with the TOA uncertainties.  We have constructed the design matrices, ${\bf M}$, and the white noise covariance matrices ${\bf C}$, for 13 of the data sets described in \citet{dfg+13}.  The design matrices encode the intricacies of the NANOGrav timing models and the non-uniform observing epochs.  Variations in TOA uncertainties are encoded in the noise covariance matrices, which we have treated as diagonal.  

\begin{figure}[tbp!]
\begin{center}
\includegraphics[height = 58mm, width = 90mm]{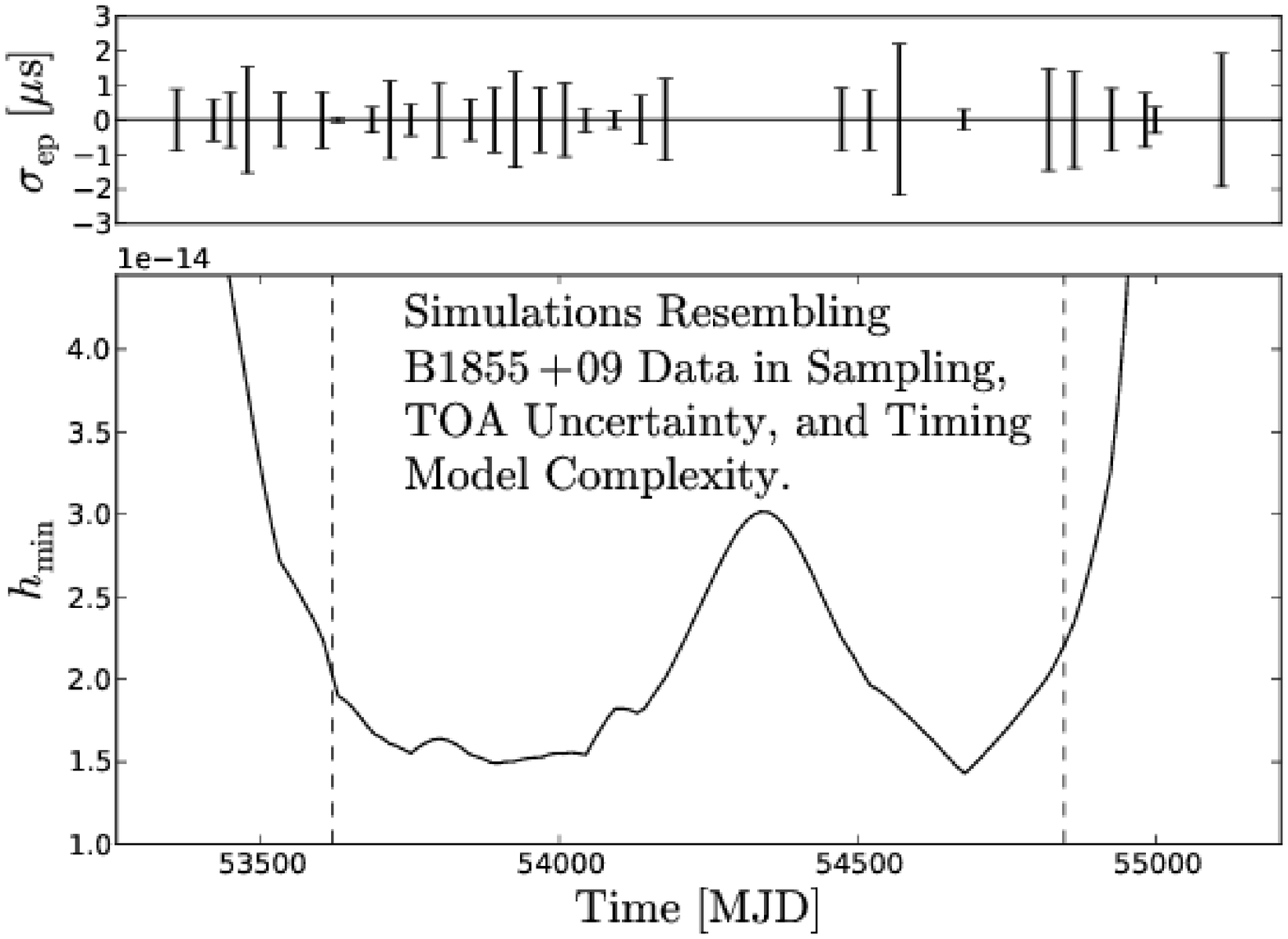}
\includegraphics[height = 58mm, width = 90mm]{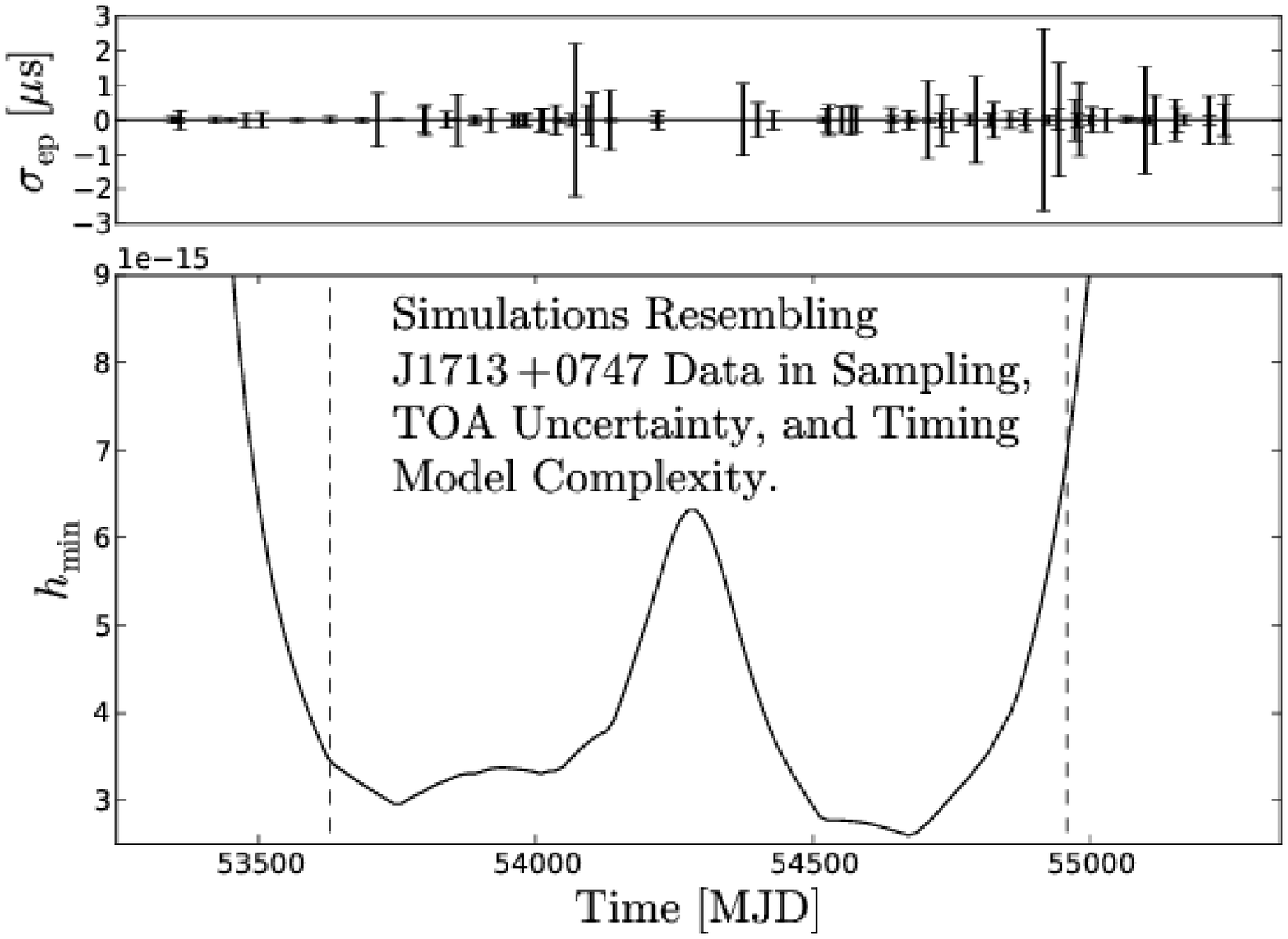}
\caption{The minimum amplitude BWM as a function of time that would appear as a 1-$\sigma$ event assuming a white noise model.  ${\rm \bf Top}$: results for simulated data resembling in sampling, TOA uncertainties, and timing model complexity the PSR B1855+09 data from \citet{dfg+13}.  Demorest et al. indicate that this pulsar has an epoch-averaged rms residual of 111~ns, very near the 100 ns rms residuals we use later in simulations of idealized pulsar timing data sets.  ${\rm \bf Bottom}$: results for similarly simulated data resembling the PSR J1713+0747 data from \citet{dfg+13}. J1713+0747 is NANOGrav's best-timed pulsar and this simulated data set was the most sensitive to BWMs of all we considered.  The top panels of these plots illustrate the TOA sampling and the epoch-to-epoch fluctuations in the TOA uncertainty in our simulated data sets. The dashed vertical lines in each plot bound the innermost 70\% of each simulated pulsar timing data set.}
\end{center}
\end{figure}

\begin{figure}
\begin{center}
\includegraphics[height=60mm,width=90mm]{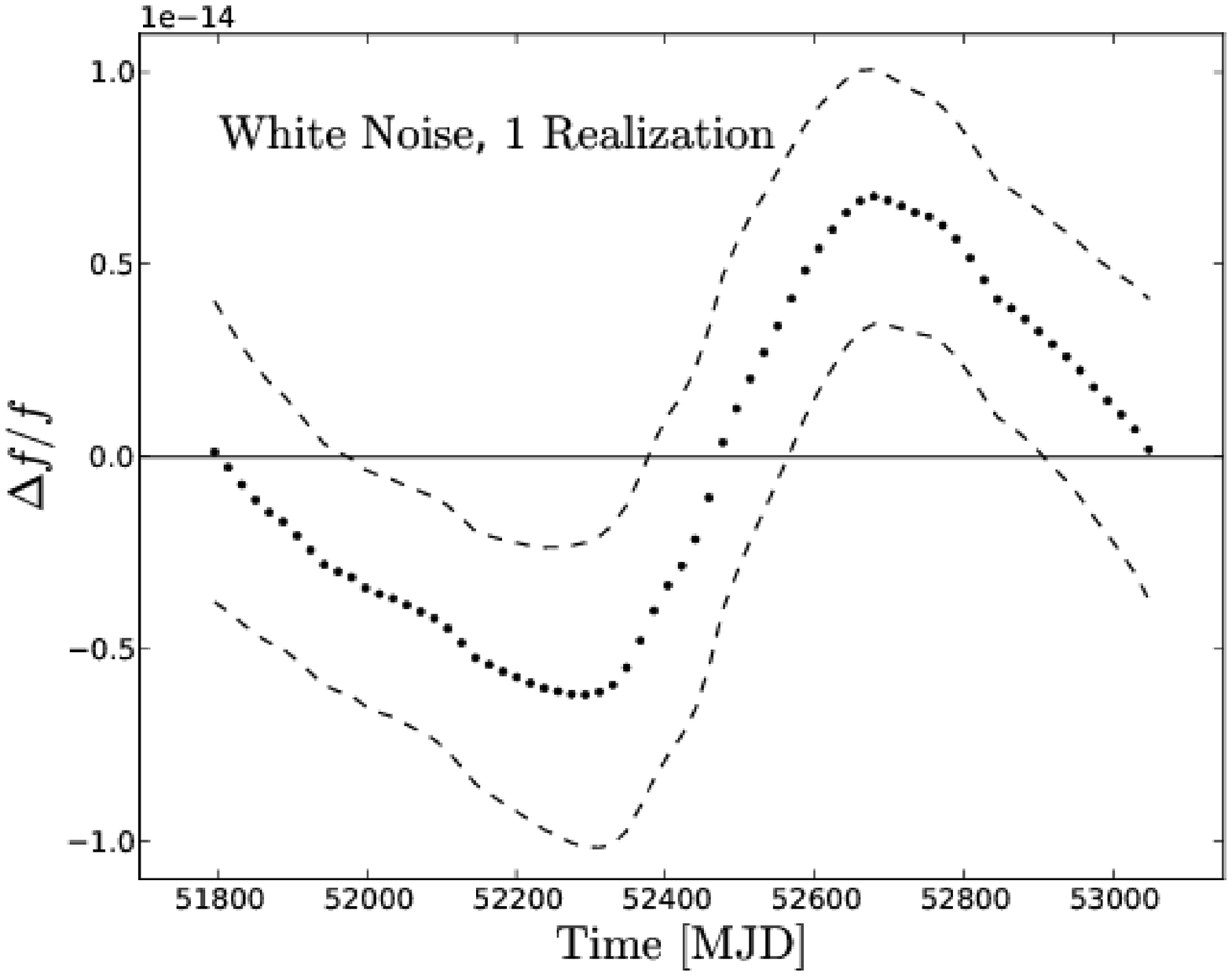}\\
\includegraphics[height=60mm,width=90mm]{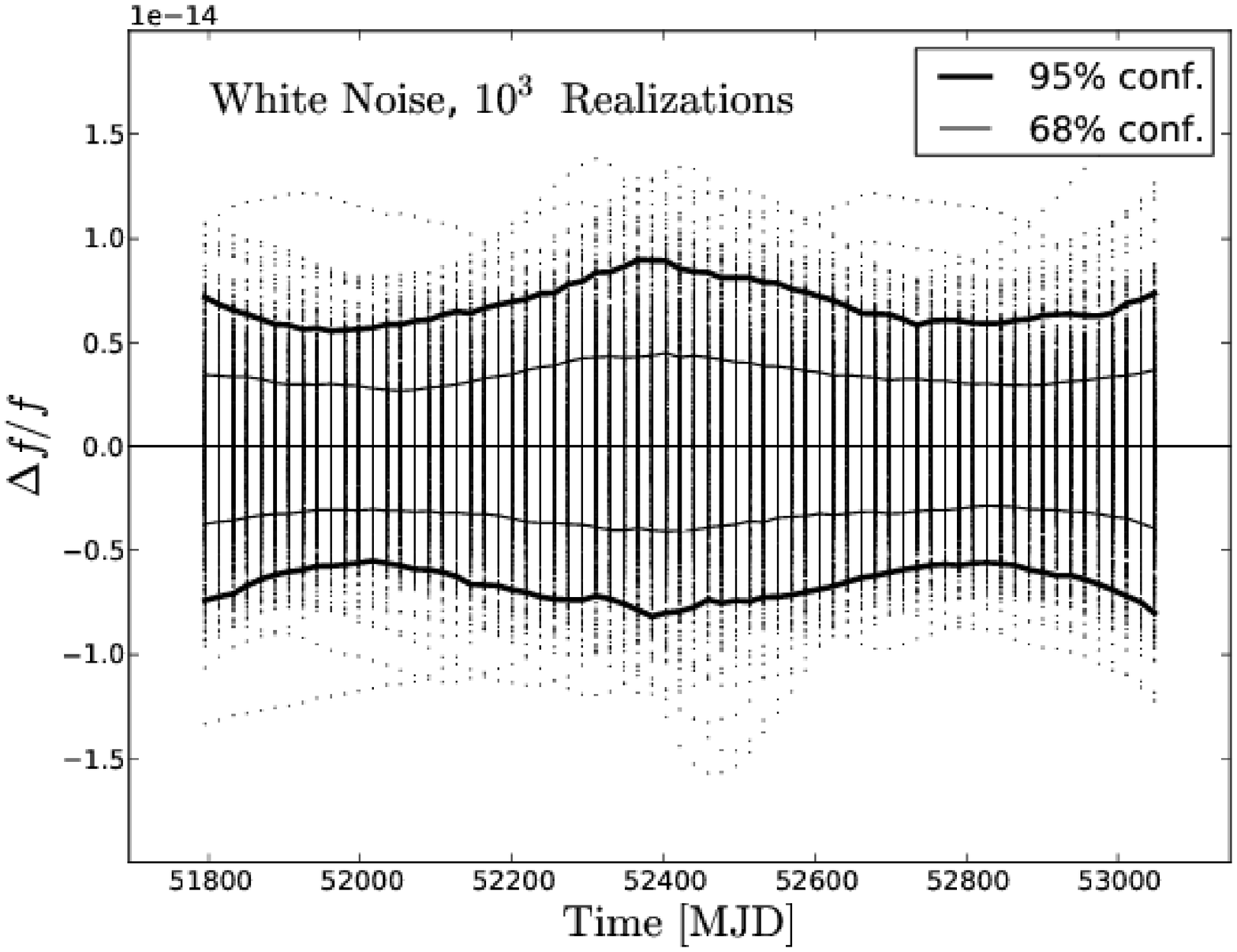}\\
\includegraphics[height = 60mm, width = 90mm]{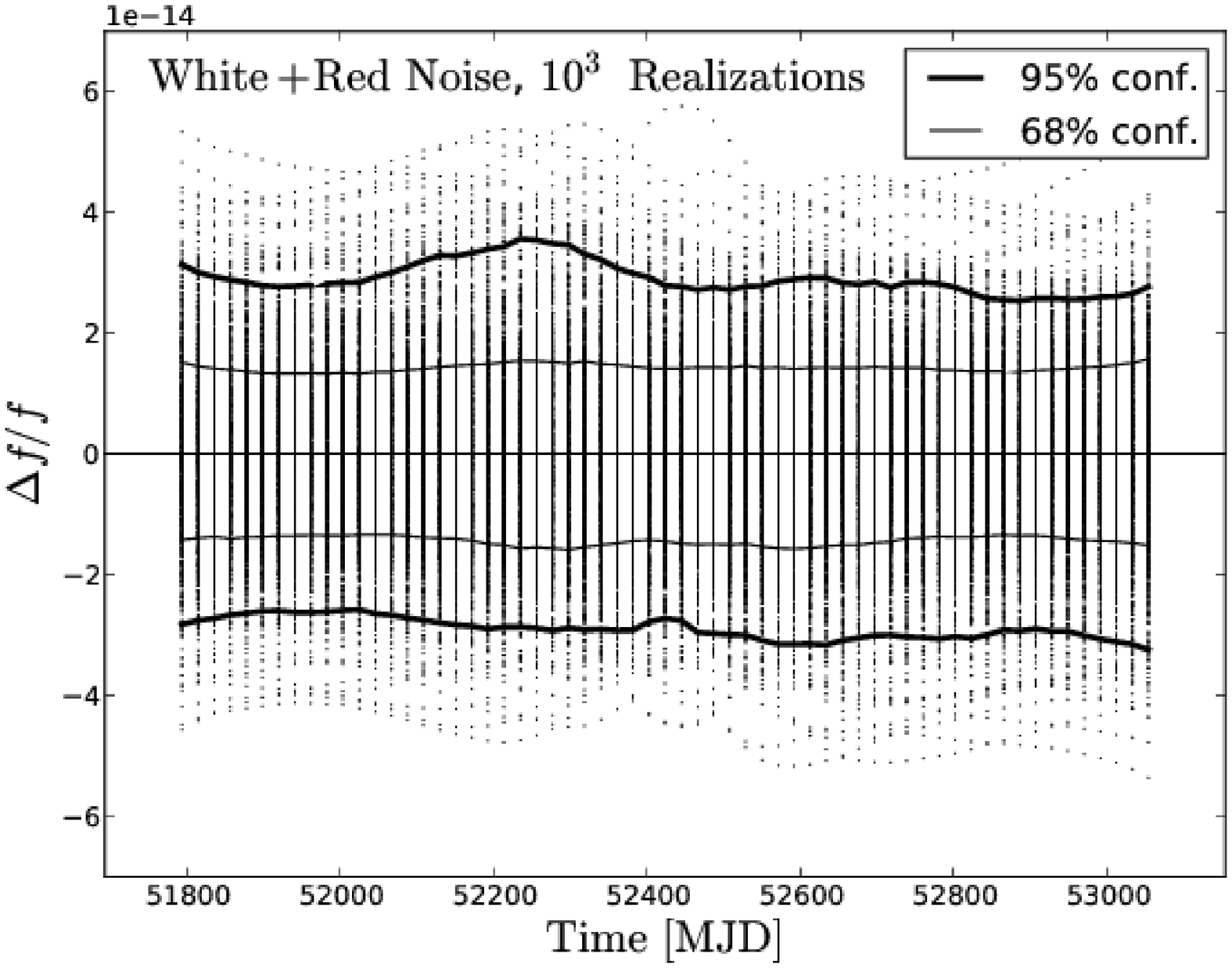}
\caption{Best-fit instantaneous fractional frequency changes as a function of trial burst time in realizations of noise-like simulated idealized timing residuals (five years, equispaced samples, uniform TOA uncertainties, only $\nu$ and $\dot{\nu}$ in timing models). ${\rm \bf Top}$: the result from a single realization of 100 ns rms white noise residuals.  The black dots indicate the best-fit values. The dashed contours indicate the 1-$\sigma$ amplitude uncertainties.  ${\rm \bf Middle}$: an ensemble of 1000 curves like the one shown in the top plot.  We have drawn contours that bound 68\% and 95\% of the points.  The 68\% contour in the middle plot agrees well with Equation (12) as can be seen in the bottom rows of Table 1.  ${\rm \bf Bottom}$: an ensemble of such curves if the residuals consist of white and red noise.  We have created 1000 realizations of simulated residuals that on average had an rms of 100 ns.  On average, half of the variance in the residuals is from white noise while half is from red noise with a power law spectrum (spectral index of 5).  Certain realizations had a smaller or larger rms than 100 ns owing to the wide variability in realizations of red noise drawn from the same distribution.}
\end{center}
\end{figure}  

\begin{deluxetable*}{cccccccccccc}[tbp]
\tablewidth{0pc}
\tablecolumns{7}
\tablecaption{BWM Sensitivity Assessment for Simulated Individual Pulsar Timing Data Sets}
\tablehead{
	\colhead{PSR}&
	\colhead{$t_0$}&
	\colhead{$t_f$}&
	\colhead{$N_{\rm TOA}$}&
	\colhead{$N_{\rm ep}$}&
	\colhead{Ep.Av.rms}&
	\colhead{$\log(h_{\rm min}^{\rm best})$}&
	\colhead{$T^{\rm best}$}&
	\colhead{$\log(h_{\rm min}^{\rm worst})$}&
	\colhead{$T^{\rm worst}$}&
	\colhead{$\log(h_{\rm min}^{\rm av})$}&
	\colhead{$\log(h_{\rm min}^{\rm rms})$}\\
	\colhead{}&
	\colhead{(MJD)}&
	\colhead{(MJD)}&
	\colhead{}&
	\colhead{}&
	\colhead{($\mu{\rm s}$)}&
	\colhead{}&
	\colhead{(MJD)}&
	\colhead{}&
	\colhead{(MJD)}&
	\colhead{}&
	\colhead{}}
\startdata
J0030+0451        &    53358    &    55107     &     545        &     21      &      0.148     &       $-$13.55      &      53744       &      $-$12.99       &       54844       &      $-$13.27      &      $-$13.77\\
J0613$-$0200    &    53448    &    55122     &     1113      &     60      &      0.178     &       $-$13.49      &      53830       &      $-$13.20       &       54279       &      $-$13.38      &      $-$14.09\\
J1012+5307        &    53217    &    55122     &     1678      &     84      &      0.276     &       $-$13.50      &      54715       &      $-$13.18       &       54271       &      $-$13.36      &      $-$14.00\\
J1455$-$3330    &    53217    &    55122     &     1100      &     73      &      0.787     &       $-$12.80       &      54405       &      $-$12.45       &      54836        &      $-$12.63      &     $-$13.31\\
J1640+2224        &    53343    &    55108     &     631        &     27      &      0.409     &       $-$13.67       &      53932       &      $-$13.15       &      54282        &      $-$12.45      &     $-$13.85\\
J1643$-$1224    &    53217     &   55122     &     1266      &     74      &      1.467     &       $-$13.45       &      53772       &      $-$13.13       &      54230        &      $-$13.34      &     $-$13.96\\
J1713+0747        &    53343    &   55245      &     2368      &     86      &      0.030     &       $-$14.59       &      54677       &      $-$14.15       &      54960        &      $-$14.42      &     $-$14.96\\
J1744$-$1134    &    53216    &    55122     &     1617      &     84      &      0.198      &       $-$13.96      &      54674       &       $-$13.43       &     53502         &      $-$13.74     &     $-$14.22\\
B1855+09            &    53358    &    55108     &     702        &     30      &      0.111      &       $-$13.94      &      54673       &       $-$13.52       &     54339         &      $-$13.71      &     $-$14.34\\
J1909$-$3744    &    53292    &    55122     &     1001      &     55      &      0.038      &       $-$14.01      &      54550       &       $-$13.59       &     53566         &      $-$13.83      &     $-$14.49\\
J1918$-$0642    &    53216    &    55122     &     1306      &     78      &      0.203      &       $-$13.39      &      54634       &       $-$12.68       &     53502         &      $-$13.10      &     $-$13.39\\
J2145$-$0750    &    53267    &    55100     &      675        &    36      &      0.202      &       $-$12.78      &      54462       &        $-$12.29      &     53542          &      $-$12.61      &     $-$13.13\\
J2317+1439        &    53358    &    55107     &      458        &    31      &      0.251      &       $-$13.69      &      54622       &        $-$13.29      &     54288          &      $-$13.54      &     $-$14.03\\\hline
Analytic                 &    51500    &    53343     &       67          &    67      &      0.100      &      $-$14.53      &       52009      &        $-$14.37       &     52422         &       $-$14.46     &      $-$15.38\\         
Sim. White            &    51500    &    53343     &       67          &    67      &      0.100      &      $-$14.54       &      52053      &        $-$14.37       &     52403         &      $-$14.47       &    $-$15.38\\
Sim. Red               &   51500    &    53343      &       67          &    67      &      0.100      &      $-$13.87       &      52025      &       $-$13.81       &      52235        &       $-$13.84      &     $-$15.20
\enddata
\tablecomments{BWM sensitivity for individual simulated pulsar timing data sets.  The top 13 rows describe the simulated NANOGrav data sets in the case of a Gaussian white noise model. These sensitivities were derived with the design matrix formalism and depend on the observing epochs, TOA uncertainties, and timing models of the data, but not the measured residuals.  The 14$^{\rm th}$ row describes the analytic sensitivity expectations from a data set satisfying the assumptions underlying Equation (12); this can be compared to the 15$^{\rm th}$ row which corresponds to the same type of idealized data set, but the values in the table come from applying our design matrix method to simulated data sets. The 16$^{\rm th}$ row correspond to results from our Monte Carlo simulations of an idealized pulsar timing data set with white and red noise.  From left to right, the rows are: the pulsar name, the MJD of the first observation, the MJD of the last observation, the total number of TOAs, the number of distinct days on which TOAs were measured, the epoch-averaged rms (taken from \citet{dfg+13}), the best 1-$\sigma$ BWM amplitude sensitivity possible with that simulated data set, the MJD at which this best sensitivity is achieved, the worst 1-$\sigma$ BWM amplitude sensitivity possible (in the innermost 70\% of the data span), the MJD at which this worst sensitivity is achieved, the average sensitivity across the innermost 70\% of the data span, and finally, the standard deviation of the BWM amplitude sensitivity across the innermost 70\% of the data span.}
\end{deluxetable*}

At each time along an equispaced grid of 100 trial burst times, $t_{Bi}$, within the innermost 70\% of each pulsar's observed time span, we have added one column to the design matrix corresponding to an additional fit parameter--the amplitude of an instantaneous change in $\nu$ at $t_{Bi}$ with no associated change in $\dot{\nu}$ and no exponential relaxation, consistent with the signature of a BWM.  We only consider the innermost 70\% of each time series because \citet{cj12} showed that in idealized pulsar timing data sets, the sensitivity to BWMs is nearly constant inside of this window, while outside of it, the sensitivity deteriorates rapidly, diverging at the boundaries (see also \citet{pbp10} and \citet{vl10}). With the additional complications we are incorporating into our simulated data sets (uneven sampling, variable data quality, and complex timing models), the BWM sensitivity will not be as nearly constant over the innermost 70\% of the data span as with idealized data or the sensitivity curves may begin to diverge noticeably inside or further outside of the innermost 70\% window than is anticipated.  However, choosing to focus on the innermost 70\% provides us with a consistent and well justified means of treating disparate data sets similarly for the purpose of comparison (see Table 1).   We then form the timing model parameter covariance matrix as in Equation (11) and isolate the diagonal element corresponding to the variance in the parameter we have introduced.  After taking the square root of this variance and normalizing it by $\nu$, we have a quantity analogous to what is described in Equation (12)--the minimum BWM amplitude necessary to appear as a 1-$\sigma$ deviation from the noise model for that pulsar's residuals at time $t_{Bi}$ so long as you assume that $B(\theta,\phi)$ is unity.

In Figure 2, we illustrate the results of this design matrix method for assessing BWM sensitivity when applied to two simulated data sets having the sampling, TOA uncertainties, and timing models of NANOGrav data sets.  The results of this procedure applied to similar simulations of all 13 NANOGrav data sets we considered are summarized in Table 1.  The method is conceptually and computationally straightforward.  It does not rely on precise TOAs, but merely on the distribution in time of observation epochs.  It is not limited to white noise models and is the appropriate technique from a maximum likelihood perspective so long as the noise is Gaussian and can be described by an appropriate covariance matrix.  

%%%%%%%%%%%%%%%%%%%%%%%%%%%%%%%%%%%%%%%%%%%%%%%%%%%%%%%%%%%%%%%%%%%%%%%%%%%%%%%%%%%%%%%%%%%%%%%%%%

\subsubsection{Assessing Pulsar Term Sensitivity with\\ Monte Carlo Simulations}
We sought to develop a more flexible method that could be used to assess the sensitivity of a timing data set to a BWM when more general noise processes are considered or when deterministic, possibly systematic effects that have not been modeled are thought to be influencing the residuals.  To do this, we have adopted a Monte Carlo approach. Using the NANOGrav timing models, we have created one thousand different sets of simulated TOAs that have the same number of TOAs, the same TOA uncertainties, and the same observing epochs (the distinct days on which TOAs were measured) as the real data sets, but have residuals consistent with white Gaussian noise.  We considered a grid of trial BWM times separated by 21 days within the innermost 70\% of each simulated time series.  At each of these trial burst times, we have again added one fit parameter to the timing model corresponding to the amplitude of sudden change in $\nu$.  We have then iteratively re-fit the timing model according to Equation~(10) until the post-fit rms and pre-fit rms of the residuals are within two nanoseconds of each other. Iteration is necessary because several of the timing model parameters, like those describing a pulsar's binary motion, influence the residuals in a non-linear fashion, contrary to the assumptions of Equation (10).  We find that the timing model with this one additional parameter converges in three or fewer iterations in all cases we considered.  

In the top plot of Figure~3, we show the results of carrying out this fitting procedure in just one realization of simulated TOAs for an idealized pulsar timing data set (equal observing cadence and TOA uncertainty, one TOA per observing epoch, and a timing model that fits only $\nu$ and $\dot{\nu}$) with 100~ns rms residuals.  For a particular realization, the best-fit frequency changes (indicated by black dots) across the observation span show correlated structure but are consistent with zero amplitude (the dashed curves indicate the 1-$\sigma$ uncertainty on the amplitude). The shape of these curves varies substantially between realizations.  When we overlay 1000 such curves, we get the result presented in the middle plot of Figure 3.  We have drawn in curves that bound 68\% (1-$\sigma$) and 95\% (2-$\sigma$) of the points.  The 68\% contour for these idealized simulations almost perfectly matches the theoretical predictions of \citet{vl10} (see Table 1 for a quantitative comparison).  We have found that this method can accurately reproduce the sensitivity curves we computed with the design matrix method (like those in Figure 2) in all the cases we considered.

In the bottom plot of Figure~3, we have carried out a procedure similar to that which produced the middle plot, but we have included both white and red noise in our simulations.  We have included red noise having a power spectrum $P(f)=A\left[1+(f/f_c)\right]^{-\alpha/2}$ where $f_c$ is a corner frequency below which the power spectrum begins to flatten (we have set this to 1/5 yr$^{-1}$; see \citet{chc+11}) and $\alpha$ is the spectral index (we have set this to 5; see \citet{sc10}).  We have chosen a value of $A$ such that over 1000 simulated sets of TOAs, the rms of the residuals is 100 ns on average with the variance split equally between white and red noise.  Different realizations have different rms values as different realizations of red noise can have dramatically different shapes.  From Figure 3, and from the bottom line of Table 1,  it is apparent that although the residuals in the red-plus-white noise case and in the white noise case have the same rms on average, the sensitivity to BWMs is greatly deteriorated by red noise; the average 1-$\sigma$ sensitivity to BWMs is worse by a factor of 4.3 compared to the case of pure white noise.

%%%%%%%%%%%%%%%%%%%%%%%%%%%%%%%%%%%%%%%%%%%%%%%%%%%%%%%%%%%%%%%%%%%%%%%%%%%%%%%%%%%%%%%%%%%%%%%%%%

\subsubsection{Realistic vs. Idealized PTAs}

We will now compare the sensitivity to BWMs in realistic timing data sets to the sensitivity of idealized timing data sets.  In the top plot of Figure 2, we plot the 1-$\sigma$ BWM sensitivity (based on our design-matrix method) for a simulated data set having the same TOA uncertainties, timing model, and number of TOAs (702) measured at the same observing epochs (the 30 distinct days on which observations were made) as the NANOGrav PSR B1855+09 data set, but consisting exclusively of Gaussian white noise-like residuals.  We show this result because the epoch-averaged residual rms quoted by \citet{dfg+13} for this pulsar (111~ns) is the closest to 100 ns which we used in our idealized simulations shown in the middle plot of Figure 3.  The epoch-averaged rms is the rms of a reduced set of effective residuals gotten by replacing all the residuals associated with a particular observing epoch with a single pseudo-residual equal to the weighted average of the residuals in that observing epoch.  In our simulations of idealized pulsar timing data sets, because we were trying to mimic the ostensible goal of \citet{dfg+13} of monthly observations over 5 years, the total number of observation epochs in this simulated B1855+09 data set (30) is smaller than we have simulated for the middle plot of Figure 3 (67) and the total length of the simulated B1855+09 data set is nearly 75 days shorter than the idealized simulations we conducted. But if we naively assume that the sensitivity scaling relation in Equation (12) ought to hold, we would expect this simulated B1855+09 data set to only have worse sensitivity compared to the idealized case by a factor of 1.7.  Instead, because of uneven sampling, fluctuating TOA uncertainty, and a more complex timing model, it is less sensitive on average than the idealized timing data set by a factor of approximately 5.6.  

To better show how the sensitivity responds to fluctuations in the observing cadence and TOA uncertainties, in the plots of Figure 2 we have included an upper panel in which we have grouped together all simulated observations that took place within one day of each other, and at that day, placed an error bar with a scale equal to the harmonic mean of the simulated TOA uncertainties from that day's observations.  For B1855+09, which was observed with Arecibo, there is a substantial gap in the observations near the middle of the data span when the dish of Arecibo was being repainted.  Though sensitivity near the middle of the observing span is anticipated to be the worst (when considering only the innermost 70\% of the observing span), in the case of our simulated B1855+09 data set with a centralized gap in its TOA coverage, it is worse than the best sensitivity by a factor of approximately 2.1 rather than 1.5 in the idealized case.

\begin{figure}
\begin{center}
\includegraphics[height=55mm,width=85mm]{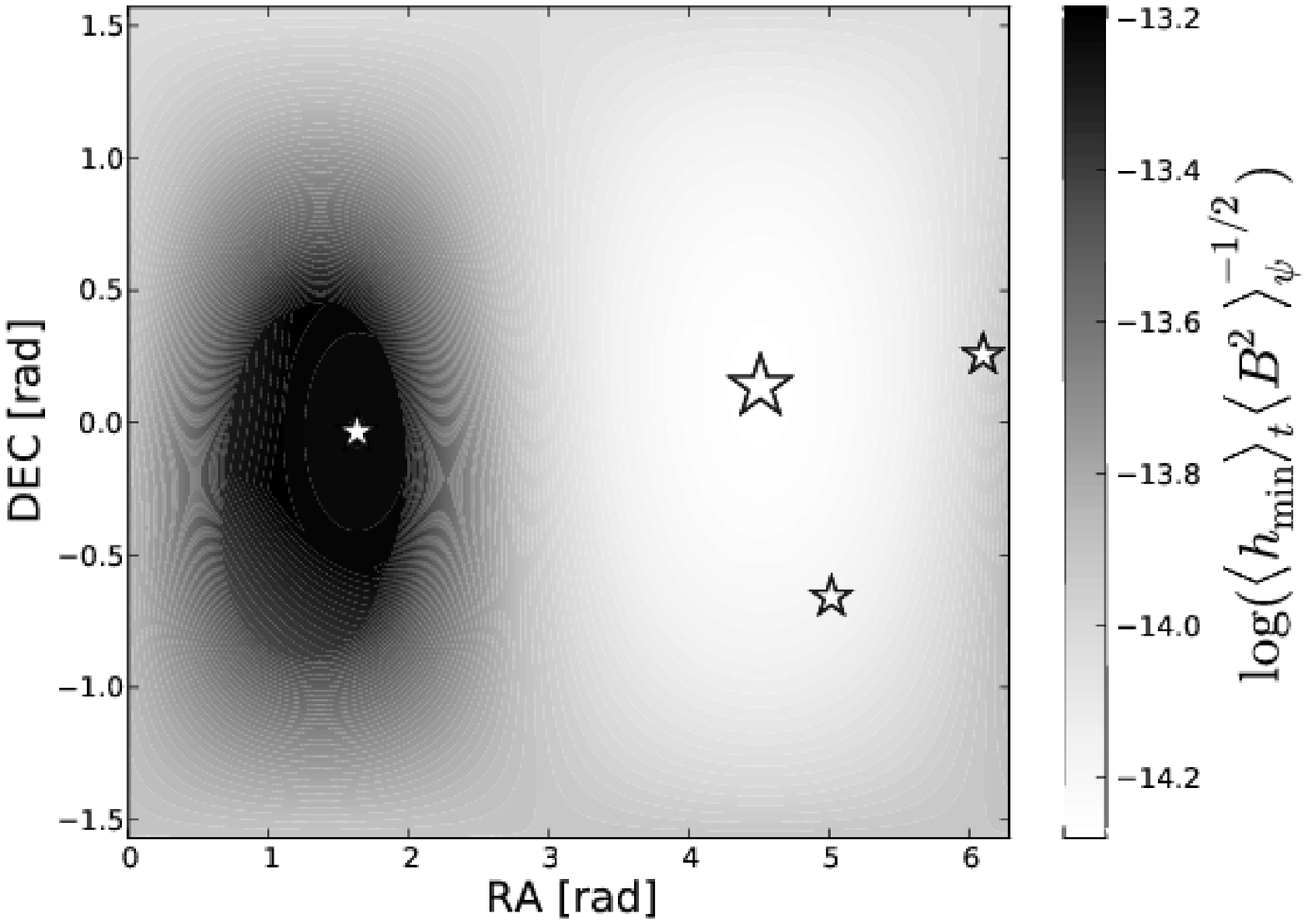}\\
\includegraphics[height=55mm,width=85mm]{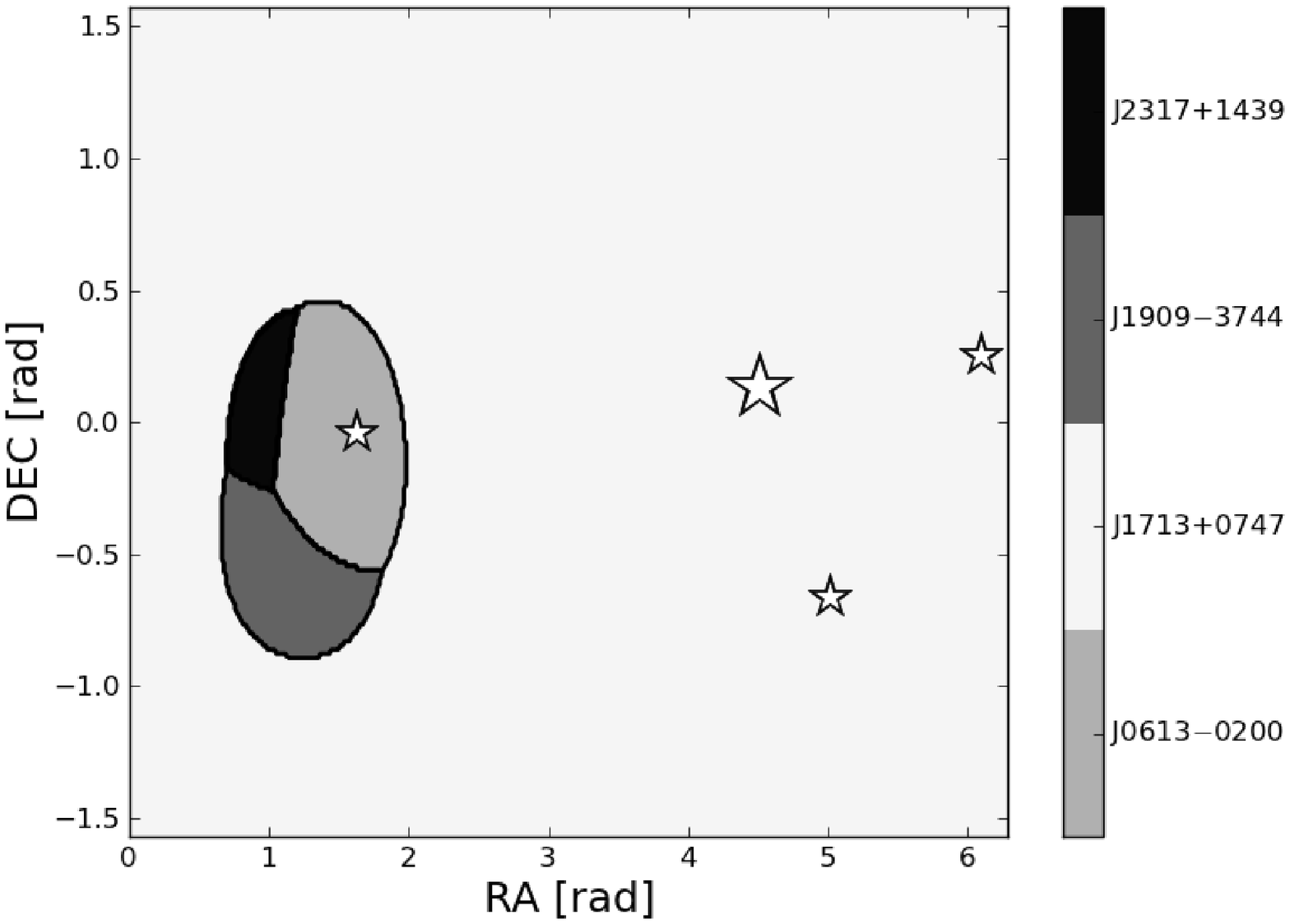}
\caption{${\rm \bf Top}$: The best time- and polarization-averaged sensitivity to BWMs from different parts of the sky from individual pulsars in our simulated PTA.  ${\rm \bf Bottom}$: indicates which simulated pulsar data set is responsible for this best sensitivity.  Our simulated data set for J1713+0747 is the most sensitive probe of BWMs unless the BWM comes from the part of the sky almost directly opposite that pulsar, and in this case, our sensitivity is much worse because other simulated pulsar data sets are either not sensitive enough or the pulsars are not located close enough to this sensitivity null to compensate.  The stars indicate the positions of the four pulsars that are most sensitive to bursts from some part of the sky according to our simulations; the biggest among them corresponds to the position of J1713+0747.}
\end{center}
\end{figure}

In the bottom plot of Figure 2, we show results based on simulations of the NANOGrav PSR J1713+0747 data set. J1713+0747 is NANOGrav's best and most thoroughly timed pulsar.  Based on the epoch-averaged residual rms quoted by \citet{dfg+13}, the number of observing epochs, and the total data span, if we again naively apply the scaling relations of Equation (12), we expect the simulated J1713+0747 data set to have better average sensitivity than our idealized pulsar simulation by a factor of 2.3.  Instead, we find that its average sensitivity is ten percent worse than the idealized data set.  Though J1713+0747 was observed with both Arecibo and the GBT, while the dish of Arecibo was being repainted, the track of the GBT had to be repaired.  So, for all of the simulated data sets we considered, those of J1713+0747 included, there is a lull in the observing cadence near the middle of the time span, and it is again apparent in the bottom plot of Figure 3 how this adversely affects the sensitivity to BWMs.  Towards the end of the trial burst times we considered, the BWM sensitivity of the simulated J1713+0747 data set diminishes dramatically; the BWM sensitivity is worst at the right edge of the window of trial burst times we have considered rather than in the middle as is anticipated in the ideal case.  

Despite the interesting features in its sensitivity curve, the simulated J1713+0747 data set is still the most sensitive to BWMs of all the data sets we simulated.  At the trial burst epoch where it is least sensitive to BWMs, the simulated J1713+0747 set is more sensitive than the data set we simulated for J1909$-$3744 is at the trial burst epoch at which it is most sensitive (J1909$-$3744 is NANOGrav's second most precisely timed pulsar).  On average, the simulated J1713+0747 data set is approximately four times more sensitive than the simulated J1909$-$3744 data set; we will discuss later the consequences of this for searching for BWMs in the Earth term.  

To this point, we have been considering the optimal case in which $B(\theta,\phi)=1$.  This is an unlikely scenario.  To illustrate the effects of the variation of $B(\theta,\phi)$ as the source direction varies, we have created Figure 4.  In it, we have taken the average sensitivities recorded in Table 1 and, based on the locations of the 13 pulsars we have considered, computed the best polarization-averaged (as in Equation (7)) single-pulsar sensitivity to BWMs from different locations on the sky.  Over most of the sky, the single-pulsar BWM sensitivity of our simulated PTA is dominated by the data set for J1713+0747, and in the parts of the sky where other pulsars become more sensitive, the sensitivity is worse by an order of magnitude than it is in the parts of the sky where the J1713+0747 data set is most sensitive.  Only four of the simulated data sets we considered are most sensitive (in a polarization-averaged sense) to BWMs from some part of the sky, and the simulated J1713+0747 data set is by far the primary player.

%%%%%%%%%%%%%%%%%%%%%%%%%%%%%%%%%%%%%%%%%%%%%%%%%%%%%%%%%%%%%%%%%%%%%%%%%%%%%%%%%%%%%%%%%%%%%%%%%

\subsubsection{Detecting and Characterizing BWMs in Pulsar Terms}

We now consider the problem of detecting and characterizing BWMs in individual pulsar timing data sets.  Because simulated J1713+0747 data sets resembling the real NANOGrav data set in sampling, TOA uncertainties, and timing model complexity are the most sensitive to BWMs of the various simulated data sets we have considered, we only consider signal detection and parameter estimation with such simulated data sets at this time.  

At an epoch $t_{\rm inj}$, we inject a BWM signal of amplitude $h_{\rm inj}$ into realizations of simulated J1713+0747 TOAs.  Without the injected signal, the simulated residuals are Gaussian white noise.  We have analyzed two different values of $t_{\rm inj}$: MJDs 54280 and 54675.  The first of these MJDs is very near the peak of the central hump in the BWM sensitivity curve depicted in Figure 2 while the second of these MJDs is very near the epoch at which the simulated J1713+0747 data set is most sensitive to BWMs.  For each of these two values of $t_{\rm inj}$, we have considered three values of $h_{\rm inj}$: $5\times10^{-15}$, $10^{-14}$, and $2\times10^{-14}$.  For each combination of $t_{\rm inj}$ and $h_{\rm inj}$, we have generated 1000 sets of simulated TOAs.

With each simulated set of TOAs, as in the Monte Carlo simulations we conducted to generate the middle and bottom plots in Figure 3, we iteratively fit for the amplitude of a BWM, $\hat{h}_B$, along a grid of trial burst times, $t_{Bi}$, using Equation (10) until the post-fit rms of the residuals is within 2 ns of the pre-fit rms.  Unlike the simulations behind Figure 3, our simulated data sets now actually contain BWMs.  The results of carrying out this procedure in one simulated data set with $t_{\rm inj}=54675$ and $h_{\rm inj}=2\times 10^{-14}$ is illustrated in Figure~5.  

In the top panel of Figure~5, we have plotted $\Gamma(t_{Bi},\hat{h}_B)=\exp{[(\chi^2_{\rm NB}-\chi^2_{\rm B}(t_{Bi},\hat{h}_B))/2]}$.  The quantity $\chi^2_{\rm NB}$ is the $\chi^2$ value of the post-fit residuals when no BWM is included in the timing model, while $\chi^2_{\rm B}(t_{Bi},\hat{h}_B)$ is the $\chi^2$ value of the post-fit residuals when a BWM at time $t_{Bi}$ is included in the timing model and the amplitude of the burst $\hat{h}_B$ is determined by typical least-squares model-fitting.  The quantity $\Gamma$ is the likelihood ratio of a timing model with a BWM at time $t_{Bi}$ compared to a model without a BWM.  In the bottom panel of Figure 5, the black diamond indicates $t_{\rm inj}$ and $h_{\rm inj}$. The values of $\hat{h}_B$ for $t_{Bi}\approx t_{\rm inj}$ are approximately equal to $h_{\rm inj}$; such large values of $\hat{h}_B$ near $t_{\rm inj}$ are approximately 7.8-$\sigma$ events as compared to the 1-$\sigma$ sensitivity curve in Figure~2, $h_{\rm min}$, and although $\hat{h}_B$ achieves highly significant amplitudes relative to $h_{\rm min}$ at values of $t_{Bi}$ remote from $t_{\rm inj}$, $\Gamma$ is significantly peaked very near $t_{\rm inj}$ ($t_{\rm inj}$ is indicated by the thick vertical black line in the top panel of Figure 5).  The small offset between the peak of the likelihood ratio and $t_{\rm inj}$ is in this case an artifact of our trial burst epoch grid spacing rather than evidence of actual bias.  Figure 5 illustrates a case where the injected BWM was detected with high significance and its amplitude and epoch were successfully recovered.  The small bump in the likelihood ratio curve near MJD 53950, an epoch situated nearly symmetrically opposite of $t_{\rm inj}$ about the mid-point of the data span, is not unique to this particular simulation, but is a generic feature of all such simulations.  For more discussion of the secondary bump in the likelihood ratio curve, see Figure 7 and the discussion related to it.

In the example in Figure 5, we claim a significant detection based on a comparison of $\hat{h}_B$ with $h_{\rm min}$, the curve depicted in Figure~2.  In keeping with this technique,  we have adopted a similar detection criterion to study our entire ensemble of simulated data sets containing injected BWMs.  We have used a time-varying amplitude threshold, $h_{\rm thresh}(\epsilon,t)=\epsilon h_{\rm min}(t)$.   At any trial burst epoch, if $|\hat{h}_B(t_{Bi})|>h_{\rm thresh}(\epsilon,t_{Bi})$, we call it a detection.  Let $N_D(\epsilon)$ be the total number of detections out of $N_{\rm sim}$ simulations.  From the Monte Carlo simulations discussed in Section 3.2.3 and depicted in the middle plot of Figure~3, we compute the number of false positive detections we expect from white noise alone for a given value of $\epsilon$, i.e. $N_{FP}(\epsilon)=N_D(\epsilon;h_{\rm inj}=0)$.  Dividing $N_{FP}(\epsilon)$ by $N_{\rm sim}=1000$, we get an estimate for the false positive fraction $f_{FP}(\epsilon)$.  By combining $f_{FP}(\epsilon)$ with the true detection fraction, $f_D=N_D(\epsilon;h_{\rm inj})/N_{\rm sim}$, we have created the receiver operator characteristic (ROC) curves in Figure~6.  We show six curves corresponding to the six $t_{\rm inj}$-$h_{\rm inj}$ pairs we considered.

\begin{figure}
\begin{center}
\includegraphics[height=75mm,width=90mm]{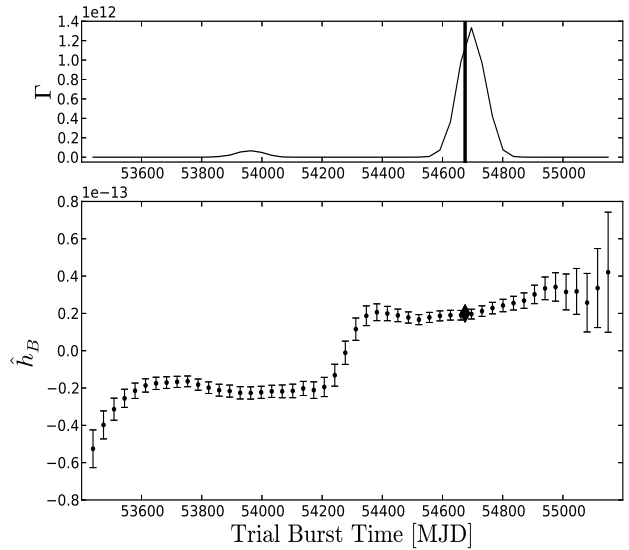}
\caption{Recovering a BWM signal injected into a simulated J1713+0747 data set.  The diamond marker indicates the epoch and amplitude of the injected burst.  By fitting for the presence of a BWM at that epoch, we are able to recover the amplitude of the injected burst.  At other epochs, the fit generates significant non-zero amplitudes for a BWM, but the likelihood ratio, $\Gamma$, for a timing model including a BWM versus one without a BWM reaches a significant global maximum very near the epoch at which the burst is injected. In the top panel, we show $\Gamma$ as it varies across the window of trial burst times.  The black vertical bar in the top panel indicates the epoch of the injected signal.  The peak of likelihood ratio is offset from the black bar by less than one grid step size.  The small bump in the $\Gamma$ curve near MJD 53950 is a generic feature of all such simulations and its amplitude relative to the other peak varies from realization to realization, sometimes becoming the primary peak.  See Figure 7 and the text pertaining to it for a further discussion of this second likelihood ratio peak.}
\end{center}
\end{figure}

\begin{figure}
\begin{center}
\includegraphics[height=70mm,width=90mm]{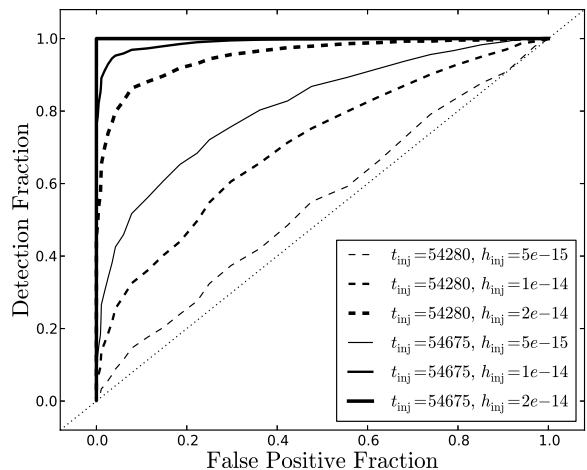}\\
\caption{The fraction of true BWMs detected as a function of the fraction of detections expected to be false and caused by Gaussian white noise.  We have implemented a simple detection scheme, looking for any trial burst time $t_{Bi}$ at which $\hat{h}_B(t_{Bi})>h_{\rm thresh}(t_{Bi})$. These are receiver operator characteristic (ROC) curves parameterized by a variable burst amplitude threshold $h_{\rm thresh}$.}  
\end{center}
\end{figure}

\begin{figure*}
\includegraphics[height=65mm,width=90mm]{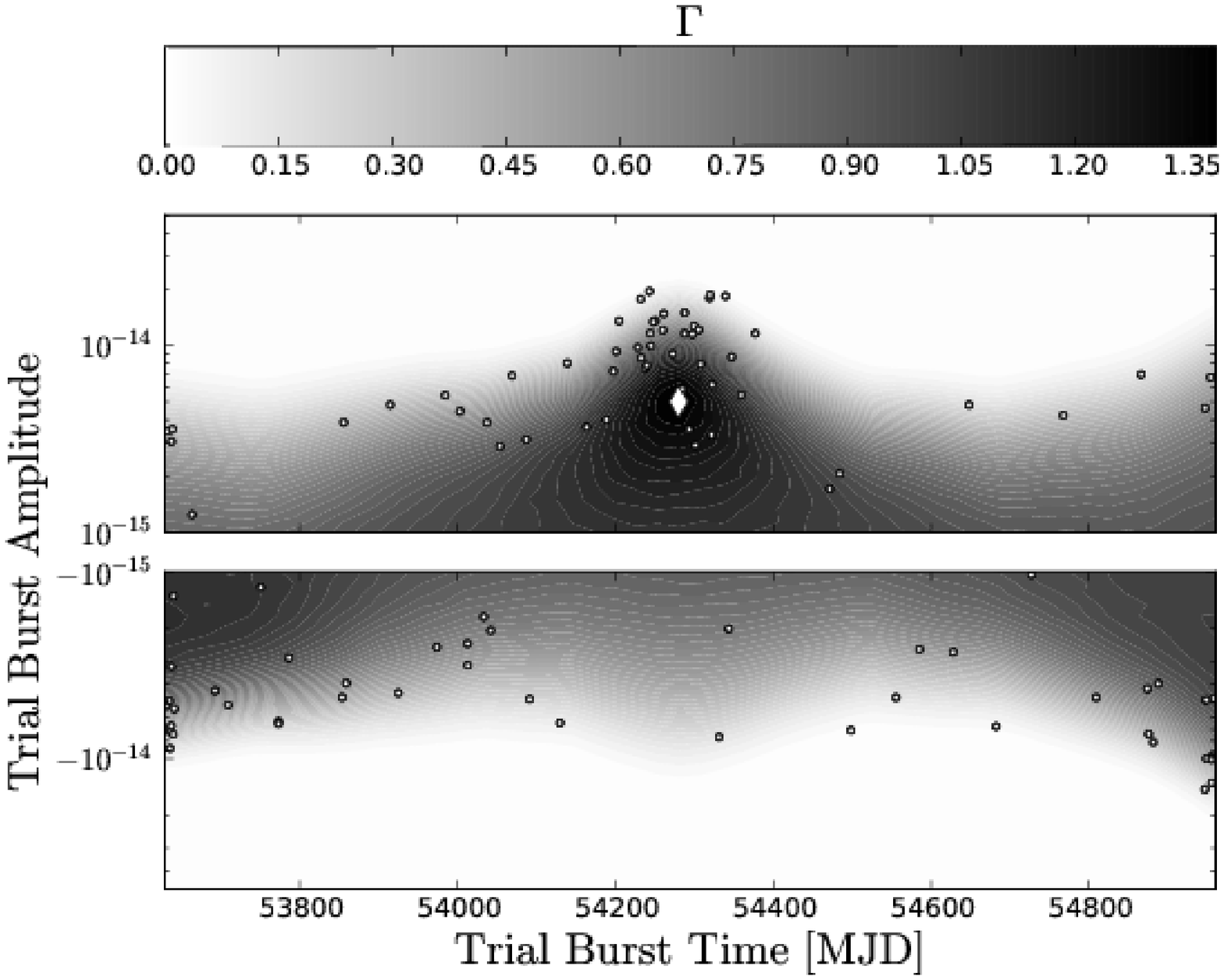}
\includegraphics[height=65mm,width=90mm]{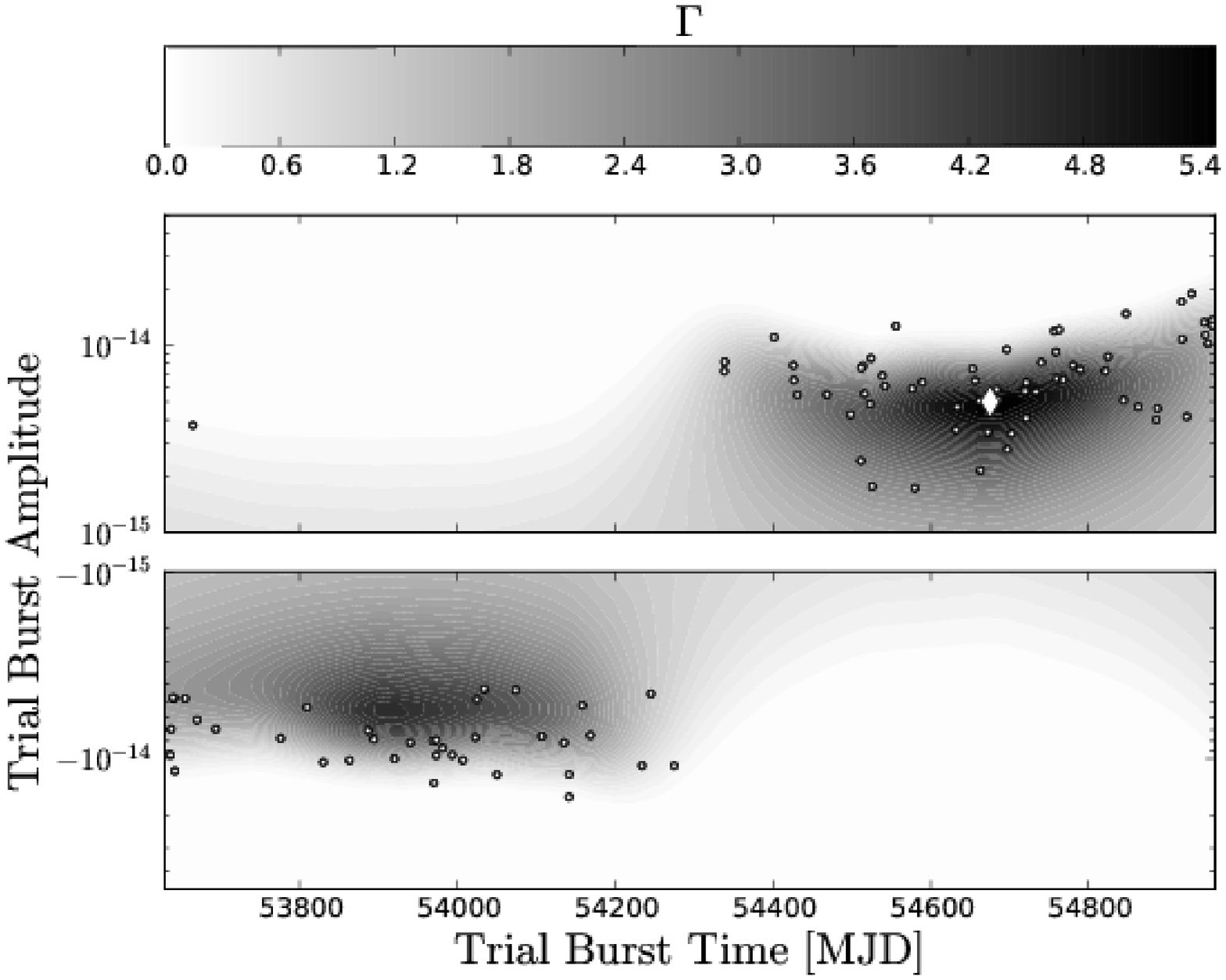}\\
\includegraphics[height=65mm,width=90mm]{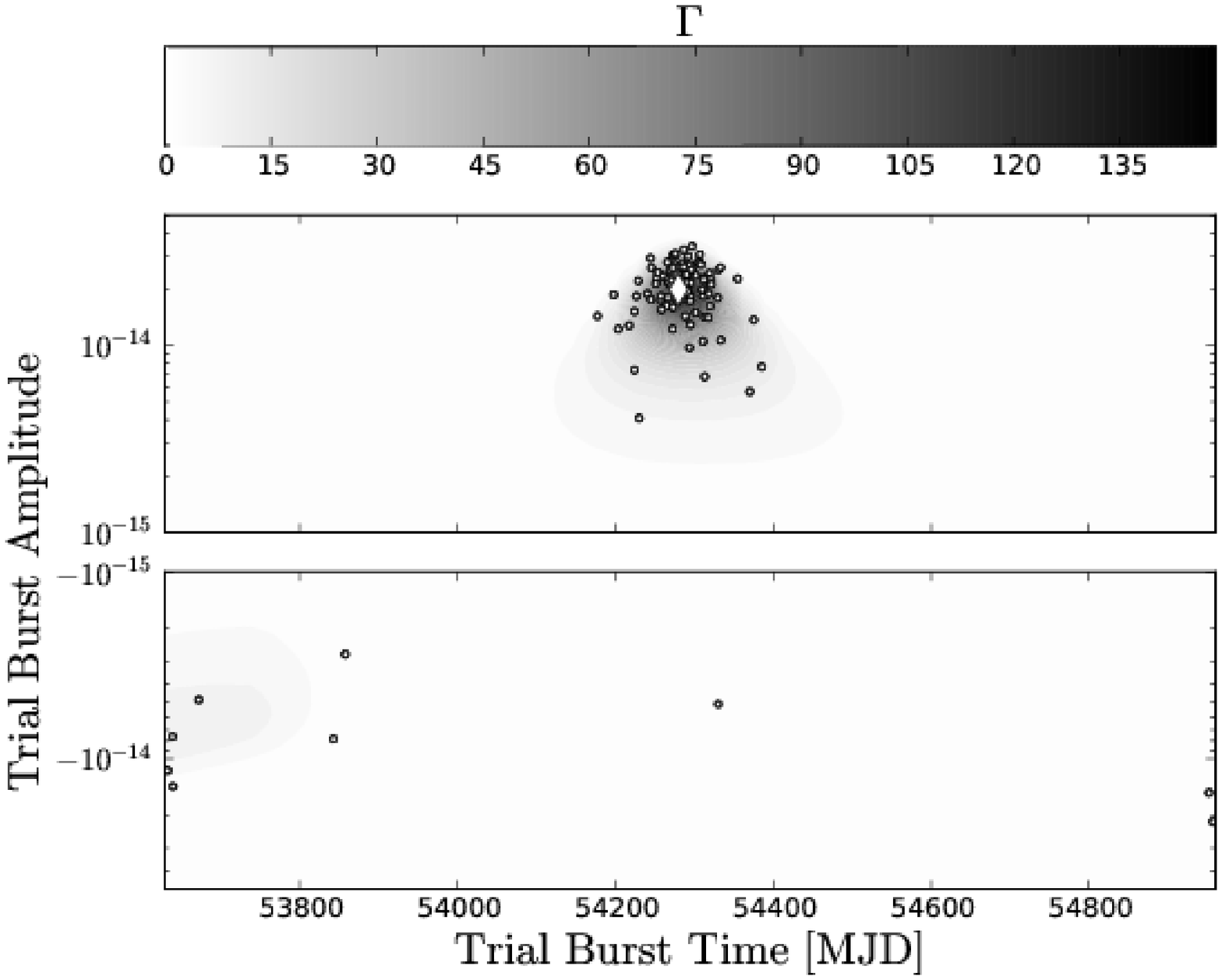}
\includegraphics[height=65mm,width=90mm]{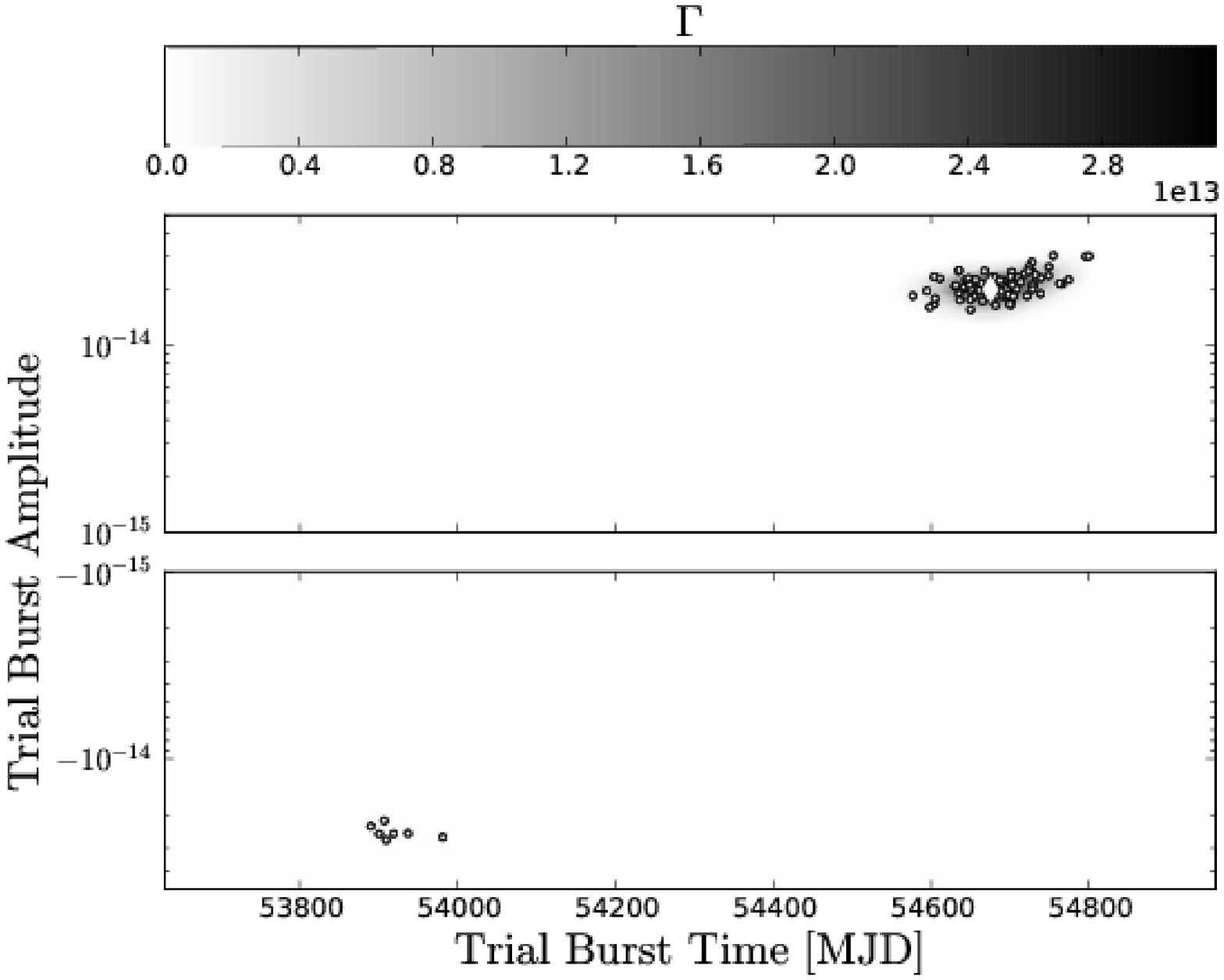}\\
\caption{Likelihood ratio surfaces for four combinations of $t_{\rm inj}$ and $h_{\rm inj}$.  For each plot, we have taken 100 simulated sets of J1713+0747 TOAs with injected BWMs and over a grid of trial burst times $t_{Bi}$ and trial burst amplitudes $h_{Bj}$, we have removed a burst with parameters described by the grid points.  We then compare the $\chi^2$ value of the post-fit residuals when we include a TOA in the timing model to the $\chi^2$ value when we do not account for a BWM in the TOAs to measure the likelihood ratio of the two models.  The color map indicates the geometric mean of the 100 likelihood ratio surfaces we computed.  The open circles indicate the peaks of the 100 different likelihood ratio surfaces.  The white diamonds indicate $t_{\rm inj}$ and $h_{\rm inj}$.  At top left, $t_{\rm inj} = 54280$ and $h_{\rm inj}= 5\times10^{-15}$.  At top right, $t_{\rm inj} = 54675$ and $h_{\rm inj}= 5\times10^{-15}$.  At bottom left, $t_{\rm inj} = 54280$ and $h_{\rm inj}= 2\times10^{-14}$.  At bottom right, $t_{\rm inj} = 54675$ and $h_{\rm inj}= 2\times10^{-14}$.  In the bottom left plot, notice the scale multiplier of $1\times10^{13}$ accompanying the color bar.}
\end{figure*}

Unsurprisingly, the ROC curves in Figure 6 show that for a fixed allowable false positive fraction and a fixed $t_{\rm inj}$, a greater fraction of brighter bursts are detected.  Also, for a fixed allowable false positive fraction and a fixed $h_{\rm inj}$ a greater fraction of BWMs are detected at $t_{\rm inj}=54675$ than at $t_{\rm inj}=54280$; this is unsurprising since MJD 54675 is approximately the epoch at which we found simulated J1713+0747 data sets to be most sensitive to BWM signatures.  With $t_{\rm inj}=54280$ and $h_{\rm inj}=5\times 10^{-15}$, this detection scheme detects BWMs at only a marginally higher rate than false detections anticipated from white Gaussian noise alone for all values of our threshold parameter $\epsilon$.  This is to be expected because at MJD 54280, $h_{\rm inj}=5\times10^{-15}$ is well below $h_{\rm min}$ so is completely consistent with white Gaussian noise.  With $t_{\rm inj}=54675$ and $h_{\rm inj}=2\times 10^{-14}$, we find that there are values of $\epsilon$ (between 3.4 and 4.8) for which all of our injected BWMs can be detected with an anticipated false alarm fraction of zero.  This is the ideal case for any signal detection scheme, though this is for the brightest bursts occurring at the most sensitive epoch we considered.  

Detection and parameter estimation are different problems.  Figure 5 depicts an ideal case in which a very bright burst occurred at an epoch at which our simulated J1713+0747 data sets were particularly sensitive and we were able to both detect it and accurately assess its amplitude and when it occurred.  This was not the case for all of our simulations.  To illustrate the strengths and weaknesses of our parameter estimation scheme, we have produced Figure 7 by considering a grid of trial burst times $t_{Bi}$ as we did in Figure 5 and in the analysis underlying Figure 6, but by also considering a grid of trial burst amplitudes $h_{Bj}$.  For the $k^{\rm th}$ of our simulations with an injected BWM described by $t_{\rm inj}$ and $h_{\rm inj}$, we remove a BWM signature described by $t_{Bi}$ and $h_{Bj}$, refit the timing model until it converges, and compute $\Gamma_k(t_{Bi},h_{Bj})$.  We do no fitting for the amplitude of a BWM.  In the plots of Figure 7, the grayscale map is of $\Gamma=\exp{[\langle\ln{(\Gamma_k)}\rangle_k]}$, the geometric mean of the likelihood ratio surfaces from all the simulated realizations we considered.  We have averaged over only 100 simulated data sets rather than 1000 because the analysis over a two-dimensional grid is more computationally expensive than the one-dimensional grid of trial burst epochs we considered before.  The white diamond in each plot indicates $t_{\rm inj}$ and $h_{\rm inj}$.  The open circles indicate the maximum values of the likelihood ratio surfaces from each of the 100 realizations we considered.

With $t_{\rm inj} = 54280$ and $h_{\rm inj} = 5\times10^{-15}$ (upper left plot of Figure~7), some peaks of the likelihood ratio surfaces do loosely cluster about the injected burst parameter values, though many are scattered broadly among both positive and negative burst amplitudes and throughout the entire window of trial burst times we considered.  The scale for $\Gamma$ is also quite low in this case, meaning that a timing model that accounts for a BWM is at best marginally favored over a timing model without a BWM included; this is again to be expected when $h_{\rm inj}$ is so small compared to $h_{\rm min}(t_{\rm inj})$.  When we keep $t_{\rm inj} = 54280$ but increase $h_{\rm inj}$ to $2\times10^{-14}$ (bottom left plot of Figure~7), the peaks of the 100 likelihood ratio surfaces we are averaging over cluster much more tightly about the injection parameters and the scale for $\Gamma$ has greatly increased, but there are still occasional spurious peaks at burst amplitudes with opposite sign from $h_{\rm inj}$ and at epochs remote from $t_{\rm inj}$.  

When we set $t_{\rm inj}=54675$, an interesting phenomenon becomes apparent.   With $h_{\rm inj}$ equal to either $5\times10^{-15}$ or $2\times10^{-14}$, there is a  tendency for some peaks of the likelihood ratio surfaces to cluster about an amplitude approximately equal to $-h_{\rm inj}$ and MJD 53950, an epoch approximately as much before the midpoint of the J1713+0747 time series as $t_{\rm inj}$ is after it.  The epoch near which the spurious likelihood ratio surface peaks in the bottom right panel of Figure 7 are clustered is the same as the epoch at which the small bump in the likelihood ratio curve occurs in the top panel of Figure 5; the secondary peak in the likelihood ratio curve in Figure 5 is not subdominant with all realizations of noise we considered.  With $t_{\rm inj}=54675$ and $h_{\rm inj}=2\times10^{-14}$, all 100 simulations we considered registered as detections with an amplitude threshold parameter $\epsilon$ large enough to completely rule out the possibility of a false positive detection (see the ROC curves of Figure 6), yet in 7 of the 100 we would have estimated the burst epoch to be approximately 700 days too early and the burst amplitude to have approximately the correct amplitude, but the wrong sign.  So, parameter estimation with this scheme is not perfect, though again, unsurprisingly, it is better for brighter bursts.   When faced with a potential detection, more sophisticated detection schemes can be developed to test the veracity of the possible detection and to improve the prospects for accurate parameter estimation.  We will discuss possible avenues for improvement in this capacity in the final section of this paper.    

%%%%%%%%%%%%%%%%%%%%%%%%%%%%%%%%%%%%%%%%%%%%%%%%%%%%%%%%%%%%%%%%%%%%%%%%%%%%%%%%%%%%%%%%%%%%%%%%%

\subsubsection{Assessing Earth Term Sensitivity}

Searching for BWMs in the Earth term, in principle, has several advantages over pulsar-term searches.  Any BWM that passes over the Earth will simultaneously begin to affect the timing residuals of all the pulsars in the PTA.  If a BWM signal is observed with high confidence to turn on simultaneously for multiple pulsars in the array, the potential that the detection is a false positive caused by some peculiarity in the behavior of a single pulsar is greatly diminished.  The magnitude of the BWM's influence on different pulsars in the array will vary as $B(\theta,\phi)$ varies from pulsar to pulsar in accordance with a specific burst polarization angle and source direction.  This allows for the recovery of the polarization of the BWM, the location of its source, and its intrinsic amplitude with some uncertainty, something that is not possible with detections in the pulsar term.  The residuals of all the pulsars in the PTA can be combined and appropriately weighted by the value of $B(\theta,\phi)$ anticipated for BWMs from a certain direction and with a certain polarization to achieve greater sensitivity to BWMs than any one pulsar in the array can attain; this is indicated by Equations (14) and (15).  In this section, we will assess the sensitivity to BWMs affecting the Earth term of the simulated NANOGrav data sets we have discussed so far and compare this performance to the performance of idealized PTAs.

To assess the potential sensitivity of NANOGrav data sets to BWMs in the Earth term, we again utilize design matrices and simulated data sets mimicking the sampling, TOA uncertainties, and timing models of the real data, but having white-noise-like residuals.  However, to accommodate parameters like the amplitude of a BWM in the Earth term which affects the timing models of all the pulsars in the array, the design matrices of each pulsar must be combined into a global design matrix.  

Suppose there are $N_P$ pulsars in the array and that the $i^{\rm th}$ pulsar has a design matrix ${\bf M}_i$ and a noise covariance matrix ${\bf C}_i$.  Without loss of generality, temporarily assume that $N_P=2$, that the first pulsar has $N_1$ timing residuals ${\bf r}_1$ at measurement epochs ${\bf t}_1$, and that the second pulsar has $N_2$ timing residuals ${\bf r}_2$ at measurement epochs ${\bf t}_2$.   For some trial choice of BWM source location, polarization angle, and epoch $t_B$, the global design matrix ${\bf M_g}$ is block diagonal except for one column corresponding to the BWM amplitude parameter:
\begin{eqnarray}
{\bf M_g} = \left[\begin{array}{c|c|c}
{\bf M}_1~ & ~{\bf 0}~ & ~B_1({\bf t}_1-t_B)\Theta({\bf t}_1-t_B)\\\hline
{\bf 0}~ & ~{\bf M_2}~ & ~B_2({\bf t}_2-t_B)\Theta({\bf t}_2-t_B)
\end{array}\right].
\end{eqnarray} 
$B_1$ and $B_2$ represent $B(\theta_1,\phi_1)$ and $B(\theta_2,\phi_2)$ respectively.  If ${\bf M}_1$ is a $n_1\times m_1$ matrix and ${\bf M}_2$ is a $n_2\times m_2$ matrix, ${\bf M_g}$ is a $(n_1+n_2)\times(m_1+m_2+1)$ matrix.  The global noise covariance matrix ${\bf C_g}$ is a $(n_1+n_2)\times(n_1+n_2)$ block diagonal matrix.  ${\bf C_g}$ and ${\bf M_g}$ can now be used to compute the parameter covariance matrix as in Equation~(11).  We take the square root of the diagonal element of the parameter covariance matrix corresponding to the amplitude of the BWM to indicate the minimum amplitude BWM that would appear as a 1-$\sigma$ deviation from the noise model.

\begin{figure}[tbp!]
\begin{center}
\includegraphics[height=55mm,width=85mm]{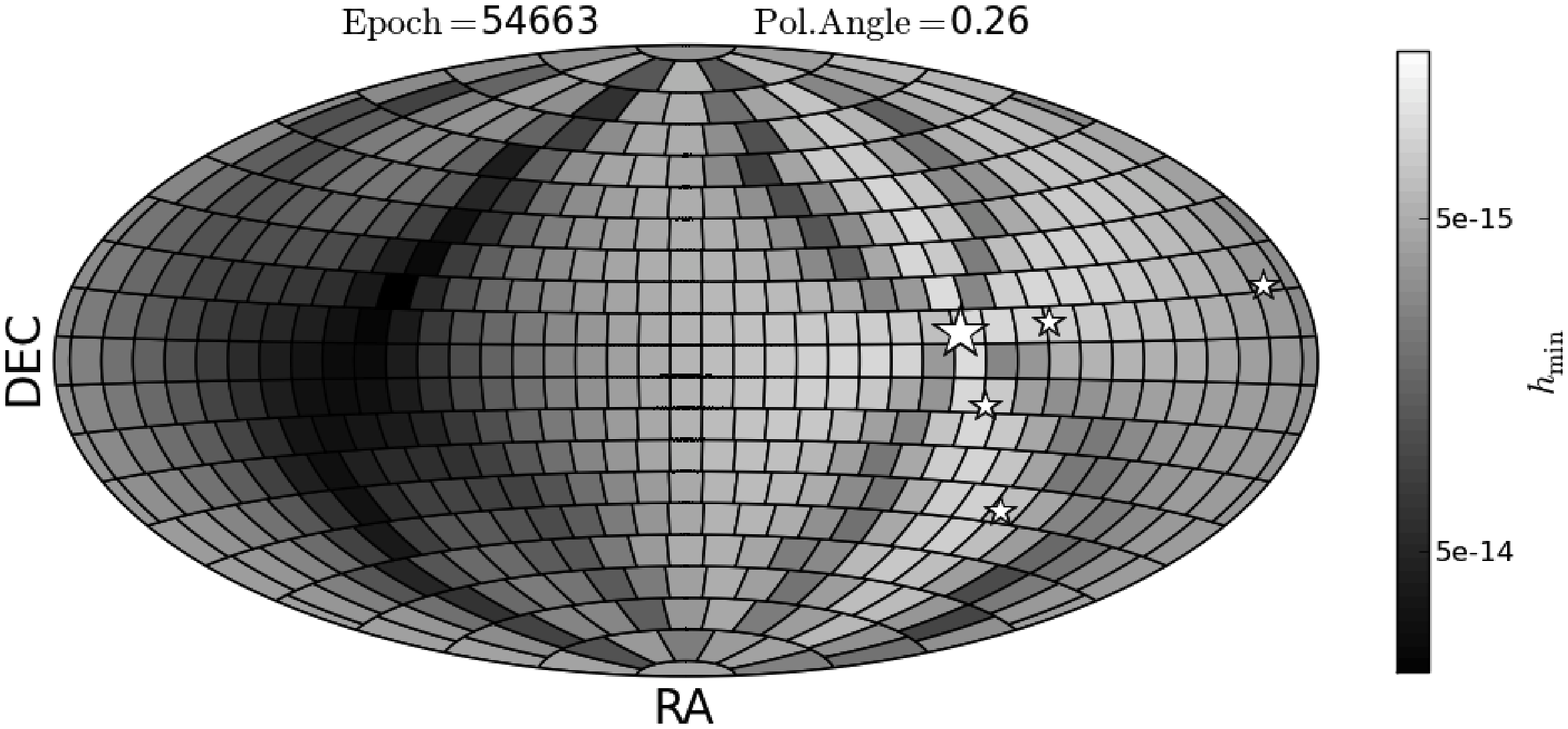}\\
\includegraphics[height=55mm,width=85mm]{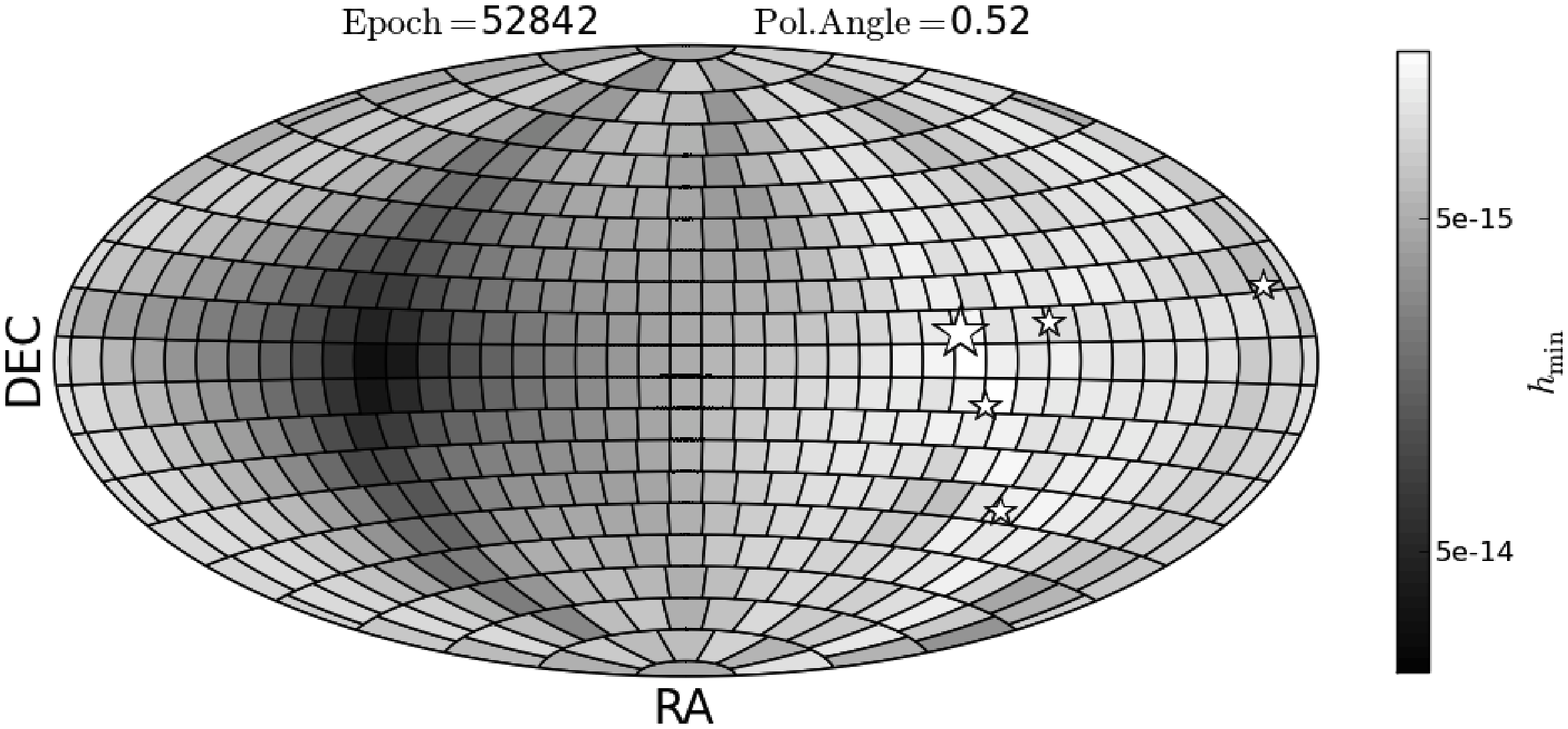}
\caption{For a specific polarization angle and burst epoch, the minimum amplitude BWM from different parts of the sky that would appear as a 1-$\sigma$ event in the Earth term given a white noise model.  ${\rm \bf Top}$: simulated PTA of 5 NANOGrav timing data sets that are most sensitive to BWMs on average according to Table~1.  ${\rm \bf Bottom}$: simulated PTA consisting of idealized pulsar timing data sets, all identical to those described in the second-to-last row of Table 1, at the locations of the 5 most sensitive NANOGrav timing data sets used to make the top plot.  The stars indicate the locations of the 5 pulsars we have considered.  The biggest star indicates the position of J1713+0747.}
\end{center}
\end{figure}

Over a four-dimensional grid of right ascension, declination, burst epoch, and polarization angle, we have used the design matrix formalism to assess the sensitivity of simulated NANOGrav data sets to BWMs in the Earth term.  Our grid consists of 520 trial source locations uniformly distributed on the sky, 50 trial burst times equispaced between MJDs 53695 and 54852, and 7 uniformly spaced polarization angles between 0 and $\pi/2$.  

We have only used the five most sensitive simulated timing data sets (according to our earlier work summarized in Table 1) to carry out the Earth-term sensitivity assessment: those associated with J1713+0747, J1909$-$3744, J1744$-$1134, B1855+09, and J2317+1439.  We initially did this to moderate the computational cost of carrying out high-dimensional global fits with many thousands of TOAs and many hundreds of fit parameters.  However, we find that the sensitivity is so thoroughly dominated by the simulated J1713+0747 data set that this simplification yields an approximately correct depiction of the simulated NANOGrav PTA's Earth-term BWM sensitivity.    For comparative purposes, over the same grid, we have assessed the sensitivity of a PTA with five ideal timing data sets like that described in the second-to-last row of Table 1 associated with pulsars at the locations of the five NANOGrav pulsars we have considered.

In Figure 8, we show the sensitivity to BWMs from these simulated data sets as it varies across the sky for a fixed epoch and polarization angle.  We show results for both the subset of five simulated NANOGrav data sets and the idealized five-pulsar PTA we have considered.  For each plot, we have selected the epoch and polarization angle that yielded the best sensitivity over the whole grid at some location on the sky.  The stars indicate the locations of the five pulsars we have included in our simulated PTA. The biggest among them represents the location of J1713+0747.  

The five-pulsar idealized PTA outperforms the five-pulsar simulated NANOGrav PTA in several important ways.  The sky-averaged sensitivity of the five-pulsar simulated NANOGrav PTA is 1.45$\times10^{-14}$ versus 6.5$\times10^{-15}$ for the idealized PTA; these numbers differ by a factor just slightly smaller than $5^{1/2}$.  If we were to compare the sky-averaged Earth-term sensitivity of the five-pulsar idealized PTA to the sky-averaged sensitivity of a single idealized timing data set, we would expect them to differ by $5^{1/2}$ (according to Equations (14) and (15)).  What we are seeing with the five-pulsar simulated NANOGrav PTA is that the sky-averaged sensitivity is entirely dominated by the simulated J1713+0747 data set--the other 4 simulated data sets do very little to improve the sensitivity--and at its most sensitivie, the simulated J1713+0747 data set just happens to be very nearly as sensitive to BWMs as one of the idealized pulsar timing data sets is at its most sensitive.  

The fact that the simulated NANOGrav five-pulsar PTA is dominated by the simulated J1713+0747 data set is further evidenced by the four lobes of relatively poor sensitivity converging on the sky location of J1713+0747.  These lobes correspond to the positions on the sky where the fixed polarization angle in this plot happens to make $B(\theta,\phi)$ exceedingly small for a pulsar at the location of J1713+0747.  In the idealized five-pulsar PTA, the other pulsars in the array are able to compensate in these regions and the sensitivity does not noticeably deteriorate.  Additionally, the region of very poor sensitivity in the part of the sky opposite J1713+0747 is not nearly as large or as insensitive in the idealized PTA compared to the more realistic five-pulsar simulated NANOGrav PTA.  This is again a consequence of the other pulsars in the idealized array being more able to compensate in regions where $B(\theta,\phi)$ becomes very small for a single pulsar.

%%%%%%%%%%%%%%%%%%%%%%%%%%%%%%%%%%%%%%%%%%%%%%%%%%%%%%%%%%%%%%%%%%%%%%%%%%%%%%%%%%%%%%%%%%%%%%%%%

\section{Constraining the Rate of BWMs}

In assessing the sensitivity of simulated individual pulsar timing data sets to BWMs in the pulsar term, we have developed a time-resolved amplitude sensitivity to BWMs from unknown locations and with unknown polarizations.  In assessing the sensitivity of combined simulated PTA data sets to BWMs in the Earth term, we have developed a time-, location-, and polarization-resolved BWM amplitude sensitivity.  Assuming a non-detection once a search for BWMs is done in real rather than simulated PTA data, we will be able to use similar sensitivity estimates to apply two distinct constraints on a quantity $\Lambda(h)$, the rate of BWMs of any polarization and from any direction arriving at a PTA with an amplitude of $h$ or greater.

If we assume that BWMs at or above an amplitude $h$ encounter a PTA as a Poisson process with a rate $\Lambda(h)$, the probability $Q$ that at least one will encounter the PTA during a period of time $\tau$ is 
\begin{eqnarray}
Q=1-e^{-\Lambda(h)\tau}.
\end{eqnarray}
Assuming that it was possible to detect a burst of at least amplitude $h$ during this time $\tau$, that none were detected, and that the rate, $\Lambda(h)$, of such bursts is known, there is a probability  $(1-Q)$ that an event did not occur by chance.  However, it may be that $\Lambda(h)$ is an overestimate.  Assuming the latter of these two possibilities, we can say 
\begin{eqnarray}
\Lambda(h)<-\frac{\ln(1-Q)}{\tau}.
\end{eqnarray}

Regarding the Earth term, we have determined that for the $i^{\rm th}$ of $N_t$ trial times in some span $T$, the $j^{\rm th}$ of $N_\Omega$ trial source locations, and the $k^{\rm th}$ of $N_\psi$ trial polarization angles between 0 and $\pi/2$, the simulated PTA had a 1-$\sigma$ BWM amplitude sensitivity $h_{{\rm min},ijk}^E$.  Consider the quantity
\begin{eqnarray}
\tau_E(h,n)=\frac{\Delta t\Delta\Omega\Delta\psi}{2\pi^2}\sum_i^{N_t}\sum_j^{N_\Omega}\sum_k^{N_\psi}\Theta(h-nh_{{\rm min},ijk}^E),\nonumber\\
\end{eqnarray}
where $\Delta t = T/N_t$, $\Delta\Omega = 4\pi/N_\Omega$, and $\Delta\psi=(\pi/2)/N_\psi$.  The quantity $\tau_E(h,n)$ is a count of the grid points in our four-dimensional search space for which an $n$-$\sigma$ detection of a BWM with an amplitude greater than $h$ is possible.  The quantity $\tau_E$ is approximately the total amount of time that the simulated PTA had $n$-$\sigma$ sensitivity to BWMs in the Earth term weighted by the fraction of the sky and the fraction of possible polarization angles over which that sensitivity was had.  This interpretation is approximate because of the finite resolution of our search grid.  

Regarding the pulsar terms, we have determined that for the $i^{\rm th}$ of $N_p$ pulsars and at the $j^{\rm th}$ of $N_i$ trial burst times within a span $T_i$ of that pulsar's simulated timing data set, there was a 1-$\sigma$ amplitude sensitivity $h_{{\rm min},ij}^P$.  Now, with a pulsar-term search, one cannot determine the BWM's source direction or polarization, but one can marginalize over all possibilities to define an equivalent quantity for the pulsar term:
\begin{eqnarray}
\tau_P(h,n) = \sum_i^{N_P}\frac{\Delta t_i}{2\pi^2}\sum_j^{N_i}\int\Theta\left(h-\frac{nh_{{\rm min},ij}^P}{B(\theta,\phi)}\right)d\psi d\Omega,\nonumber\\
\end{eqnarray}
where $\Delta t_i = T_i/N_i$. The quantity $\tau_P(h,n)$ is approximately the total amount of time that our simulated PTA had $n$-$\sigma$ single-pulsar sensitivity to BWMs with amplitudes greater than $h$ coming from any part of the sky and with any polarization. Again, this interpretation is approximate because of the finite resolution of our search grid, which in the case of a pulsar-term search, is only a one-dimensional grid over burst epoch.  The quantity $\tau_P(h,n)$ will exceed the total length of the PTA observing campaign for certain values of $h$ and $n$ because the pulsar term for each pulsar in the array is causally distinct from the pulsar terms of any other pulsars in the array. Combining Equations (18)-(20), we can apply distinct pulsar-term- and Earth-term-based constraints on $\Lambda(h)$, namely $\Lambda_{E,P}(h,n,Q)<-\ln{(1-Q)}/\tau_{E,P}(h,n)$.  

\begin{figure}
\begin{center}
\includegraphics[height=70mm,width=90mm]{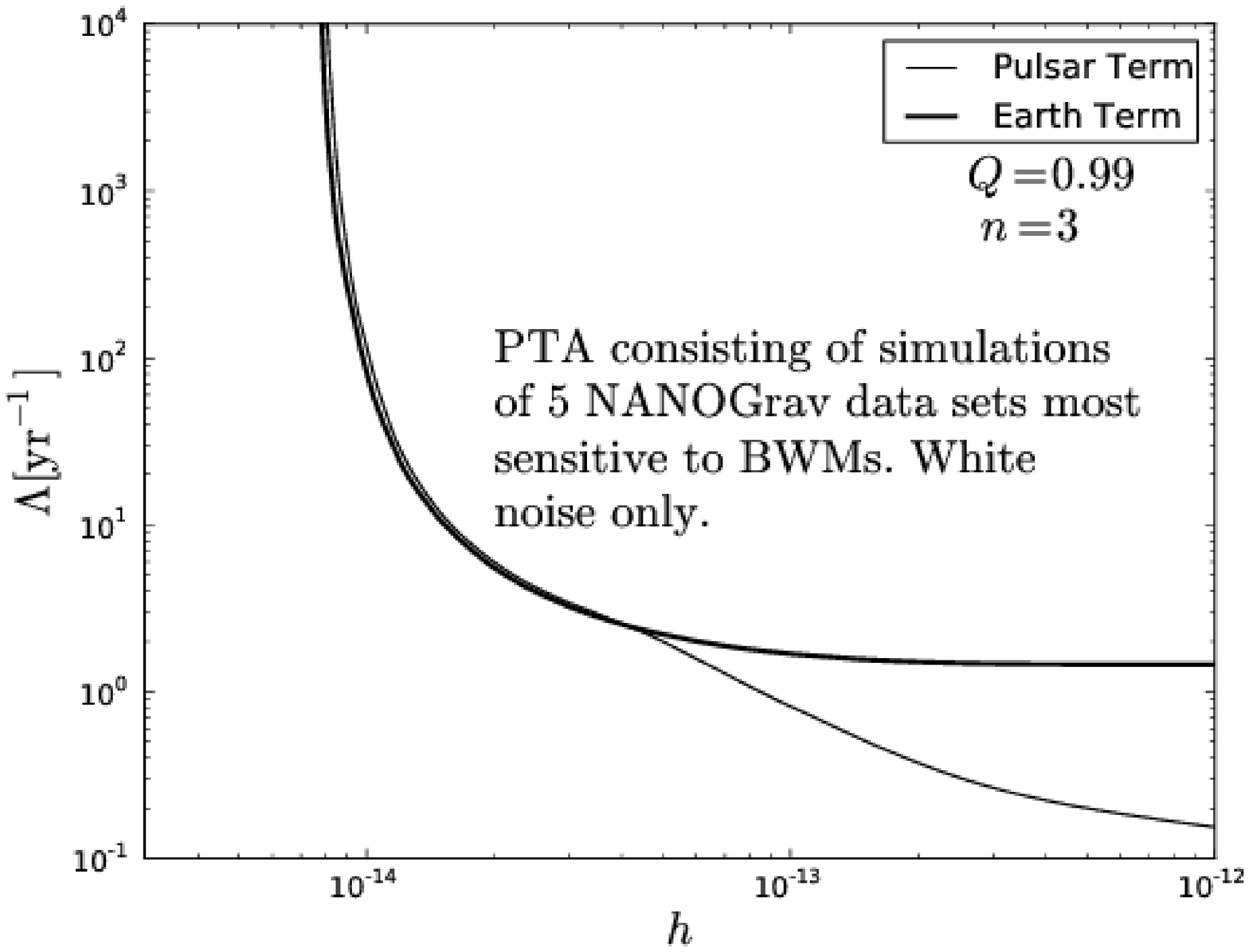}\\
\includegraphics[height=70mm,width=90mm]{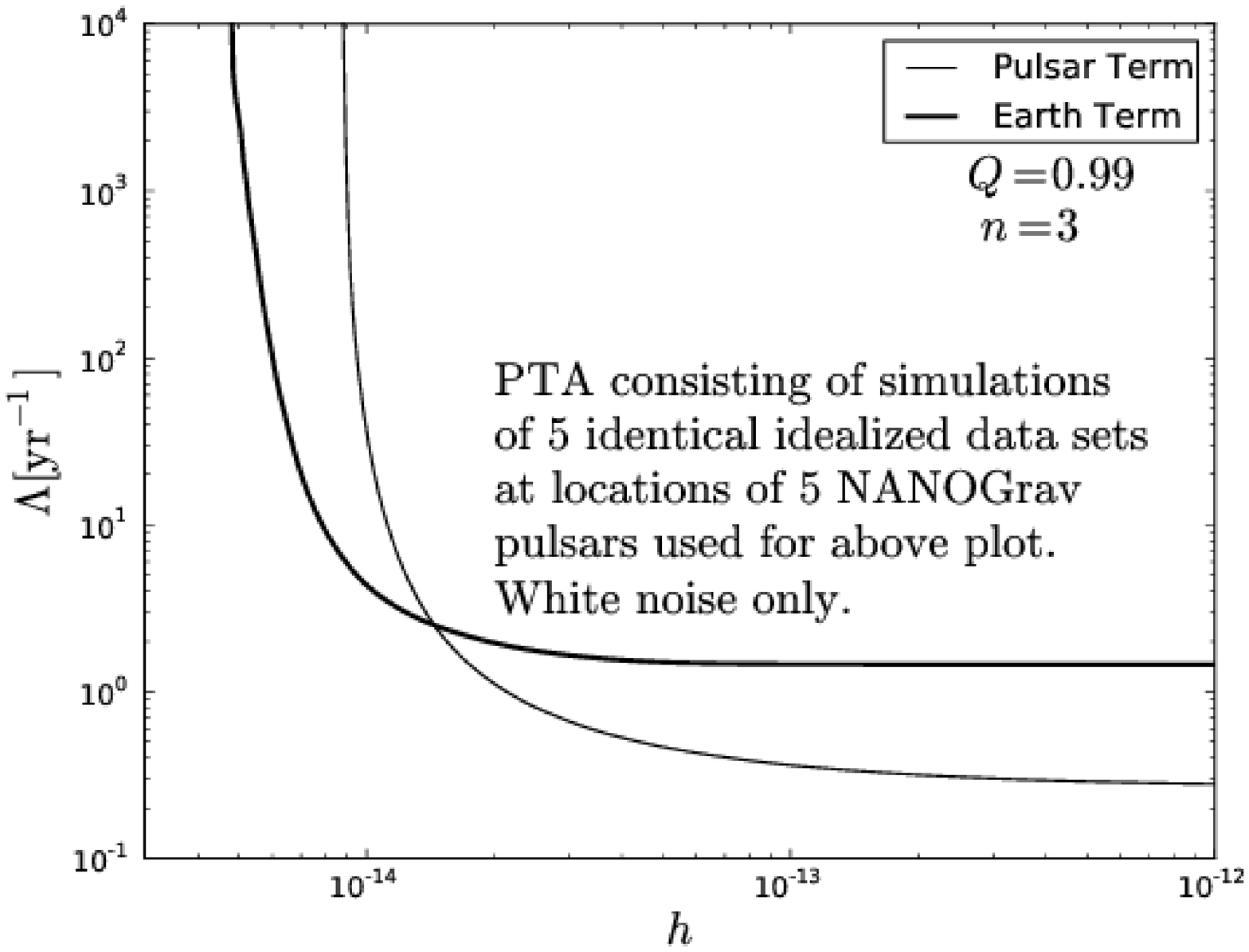}
\caption{Distinct pulsar term and Earth term constraints on the rate of BWMs above a certain amplitude assuming a non-detection in simulated PTAs.  In the top plot, we have considered only simulations of the five NANOGrav data sets most sensitive to BWMs.  In the bottom plot, we have considered an idealized PTA of five fake, identical pulsars (100 ns rms, uniform TOA uncertainty and observing cadence, basic timing model) with the same locations as the NANOGrav pulsars we have considered. We have simulated only white noise.}
\end{center}
\end{figure}

In Figure 9, we plot and compare the constraints on $\Lambda(h)$ that are possible with a PTA consisting of only the five simulated NANOGrav data sets most sensitive to BWMs and a PTA consisting of five identical idealized timing data sets associated with pulsars at the same locations as the NANOGrav pulsars we have considered.  We emphasize that these are the most optimistic constraints feasible with these simulated PTAs as they are based on white noise models.  For high values of $h$, these curves flatten out to a small value of $\Lambda$.  For the Earth-term constraint, the curve approaches a value proportional to $T^{-1}$ where $T$ is the total number of years that the timing data sets of the PTA overlap and have been searched for a BWM.  The value of $h$ above which the Earth-term constraint is flat is set by the Earth-term sensitivity of the PTA at its least sensitive.   BWMs above this amplitude would have been detected if they had occurred anywhere on the sky, at any time, and with any polarization.    For the pulsar-term constraint, the curve approaches a smaller value of $\Lambda$ proportional to $(\sum_iT_i)^{-1}$ where $T_i$ is the number of years in which the $i^{\rm th}$ pulsar's timing residuals have been searched for a BWM.  Since all of the pulsars in the array are different distances from the Earth and in different locations on the sky, the pulsar terms are probing distinct periods in the past, so the total time baseline is the sum of the individual baselines.  

For large $\Lambda$, the Earth- and pulsar-term constraints asymptotically approach small values of $h$ that are set by the optimal sensitivity to BWMs capable with Earth- or pulsar-term searches respectively; call them $h^{E}_\infty$ and $h^{P}_\infty$ .  The best sensitivity achievable with the most sensitive single pulsar timing data set in the array sets $h^{P}_\infty$.  The amount by which $h^{E}_\infty$ is less than $h^{P}_\infty$ is a measure of how much the optimal amplitude sensitivity of the array is enhanced by combining the residuals of all the pulsars in the array and fitting for the amplitude of a BWM as a global parameter.  In the bottom plot of Figure 9 associated with our analysis of an idealized PTA of identically sensitive pulsars, $h^{E}_\infty/h^{P}_\infty=0.54$--very nearly a factor of two enhancement in sensitivity is gained by combining the residuals of all the pulsars in the PTA.  With the five best simulated NANOGrav data sets we have considered, this factor is much more modest--$h^{E}_\infty/h^{P}_\infty=0.96$.  Much like the top plot of Figure 8, this further indicates that the Earth-term sensitivity of this simulated NANOGrav PTA is only marginally enhanced by the consideration of pulsar data sets other than the one for J1713+0747.

%%%%%%%%%%%%%%%%%%%%%%%%%%%%%%%%%%%%%%%%%%%%%%%%%%%%%%%%%%%%%%%%%%%%%%%%%%%%%%%%%

\subsection{Relating BWM Constraints to the\\SMBHB Population}

Constraints on the rates of BWMs of various amplitudes provide information regarding the population of SMBHBs and their evolution.  Conversely, inferences regarding the population of SMBHBs can yield predictions about the expected rate of BWMs.  Following \citet{cj12} and \citet{s13}, the rate of BWMs above a certain amplitude can be written as
\begin{eqnarray}
\Lambda(h)&=&4\pi D_H^3R(0)\int_0^{\infty}\!\!\! dM_1\int_0^{1/4}\!\!\! d\eta\int_0^{s(h)} dxx^2\frac{\Phi(M_1,z)}{M_1\ln{10}}\nonumber\\&&~~~~~~~~~~~~~~~~~~~~~\times{\cal F}(M_1,\eta,z)\frac{R(z)/R(0)}{1+z}.
\end{eqnarray}
Here, $z$ is the redshift at comoving distance $D$, $\eta=\mu/M$ is the symmetric mass ratio, $D_H=cH_0^{-1}$, $R(z)$ is the SMBHB merger rate per unit comoving volume, $\Phi(M_1,z)=dn/d\log{M_1}$ is the mass function for BHs of mass $M_1$, and ${\cal F}(M_1,\eta,z)$ is the fraction of BHs with mass $M_1$ that are paired with another BH to form a SMBHB with symmetric mass ratio $\eta$.  The function $s(h)=C_h\eta M/D_Hh$ where the amplitude of a BWM can be expressed as $h_B=C_h\eta M/D$.  For any particular binary, $C_h$ depends on the inclination angle and the angular momenta of the BHs in the binary.

\citet{s13} introduced $\Phi$ and ${\cal F}$ in a discussion of the SB of GWs, arguing that these functions can be inferred from observations of galaxies if the observations are partnered with some program for connecting the properties of the SMBHs to host galaxies, i.e. through stellar bulge mass \citep{hr04}, stellar velocity dispersion \citep{bcc+12}, or mid-infrared luminosity \citep{s11}.  The amplitude of the SB and the rate of BWMs are inextricably related.   Building on this deep connection, \citet{cj12} show in detail how amplitude constraints on the SB can constrain the rate of BWMs.  Using their Equation (22), we can say
\begin{eqnarray}
\Lambda(h)<0.32~{\rm events~yr}^{-1}A^2_{\rm lim,-14}~~~~~~~~~~~~~~\nonumber\\\times\left(\frac{{\cal M}_{c,{\rm nom}}}{{\cal M}_c}\right)^{5/3}\frac{\langle J_\Lambda(h)\rangle_{M,\eta}/10^{-3}}{J_{h_c}},
\end{eqnarray}
with a rms dimensionless SB strain $h_c(f)=(10^{-14}A_{\rm lim,-14})f^{-2/3}$. The quantity ${\cal M}_c=\langle M_c^{5/3}\rangle^{3/5}_m$ is a chirp mass characteristic of the population of SMBHBs. They have chosen a nominal SMBHB chirp mass ${\cal M}_{c,{\rm nom}}=2.3~\times~10^7~M_{\odot}$. The quantity $\langle J_\Lambda(h)\rangle_{M,\eta}$ is the average rate of bursts with amplitudes greater than $h$ over all possible $M$ and $\eta$.  The authors state that $\langle J_\Lambda(h)\rangle_{M,\eta}$ can be as large as unity, but in one Monte Carlo simulation, they computed it as $3\times 10^{-4}$ (with a detection threshold of $h=10^{-15}$).  They cite \citet{jb03} in asserting that $J_{h_c}$ can range between 0.1 and 100.  

The parameters in Equation (22) have wide ranges of possible values owing to poor constraints on $\Phi$, ${\cal F}$, and $R$ in Equation (21). \citet{s13} has considered a vast array of possible galaxy assembly histories and SMBH mass functions and finds that with 99.7\% confidence, $1.1\times10^{-16}<A<6.3\times10^{-15}$.  The current most stringent experimental 95\% confidence upper bound is $A<2.7\times10^{-15}$ \citep{src+13}.  Combining the lower bound of Sesana, the upper bound of Shannon et al., and the range of parameters in Equation (22) discussed by Cordes \& Jenet, with ${\cal M}_c={\cal M}_{c,{\rm nom}}$, we can say $1.2\times 10^{-5}<\Lambda(10^{-15})<2.3$ where $\Lambda$ is measured in events yr$^{-1}$.
%%%%%%%%%%%%%%%%%%%%%%%%%%%%%%%%%%%%%%%%%%%%%%%%%%%%%%%%%%%%%%%%%%%%%%%%%%%%%%%%%%%%%%%%%%%%%%%%%

\section{Discussion}

In this paper, we have developed general methods for assessing the sensitivity of PTA data sets to GW BWMs.  Our methods are applicable in cases with uneven observing cadence, variable TOA uncertainties, and elaborate timing models.  We have mainly considered white noise, but our methods are more widely applicable.  Through manipulations of the timing model design matrix, we can assess the PTA's BWM sensitivity whenever the noise is Gaussian and well described by a covariance matrix.  With Monte Carlo simulations, we have addressed how BWM sensitivity in the presence of more general noise processes or various unaccounted for deterministic processes can be assessed; we have used this method to demonstrate how red timing noise can prove detrimental to PTA BWM sensitivity.  

Our Monte Carlo approach for assessing PTA BWM sensitivity relies on a basic procedure for searching for BWM signatures in timing data.  By injecting BWM signals into simulated timing data, we have shown that this search procedure can reliably detect bright bursts, but that burst parameter estimation with this scheme can be biased in some instances.  Our future work on this topic will involve developing improved procedures for searching for, characterizing, and constraining BWMs in timing data sets.  Developing appropriate noise models will be a crucial ingredient to this.  \citet{chc+11} have developed a spectral method for estimating the noise covariance matrix which, along with Cholesky whitening, reduces bias in timing model parameter estimation, even in the presence of correlated noise.  However, \citet{vl13} have shown that Bayesian timing analysis can produce results which, though largely comparable to the results derived with frequentist Cholesky whitening schemes, are less biased in estimates of $\nu$ and $\dot{\nu}$, the timing model parameters most covariant with the signature of a BWM.  Such Bayesian techniques may thus prove invaluable to a proper search for BWMs \citep{vv13,lah+14}.  \citet{vv13} espouse the virtues of Bayesian pulsar timing analysis and make several key points.  Bayesian techniques allow for the incorporation of prior probability distributions for timing model or noise model parameters.  Bayesian methodologies allow for natural model comparison through marginal likelihoods; this is not unlike F-tests in frequentist analysis, but is more general.  With Bayesian techniques, joint probability distributions can be naturally computed for timing model parameters and noise model parameters so that the evidence for BWM detection can be assessed for a range of potential noise models.  Finally, in the low S/N regime in which any initial detection of a BWM is likely to reside, Bayesian inference can reliably estimate timing model parameter uncertainties while accounting for non-linear dependencies in the timing model in a way that least-squares fitting procedures cannot. 

We have found through simulations that the NANOGrav PTA's sensitivity to BWMs will likely be dominated by the data for J1713+0747, but this may change slightly with more realistic noise modeling.  Such weight assigned to a single pulsar undermines many of the benefits of a search for BWMs in the Earth term.  The sensitivity enhancement gained by coherently combining the data from many or all of the pulsars in the array is very small.  Furthermore, the prospect of discriminating a BWM from phenomena intrinsic to a single pulsar by detecting it simultaneously in more than one pulsar is only possible for bright BWMs visible in the residuals of the second- or third-most sensitive timing data sets in the array.  We suspect that NANOGrav is not the only PTA with a single dominant pulsar timing data set when it comes to BWM sensitivity.  For instance, in its 10 cm timing data, the PPTA has 475 TOAs with an rms residual on one-year timescales of 58 ns for J0437$-$4715.  By these two measures, their next-best timing data set is for J1909$-$3744 with a modestly larger residual rms on one-year timescales (83 ns) and only 138 TOAs \citep{mhb+13}.  Though these quantities do not completely determine the BWM sensitivity, we believe that they indicate that the PPTA's sensitivity will be dominated by their J0437$-$4715 data set.  

To benefit Earth-term searches for BWMs, PTAs should work to enhance the quality of their second- and third-best timing data sets to try and bring them in line with their best data sets, whether this be through an increased observing cadence or increased integration times per TOA measurement.  However, as the amount of available telescope time is finite, we recognize that this observing strategy is in tension with proposed optimal observing strategies for detecting a SB of GWs \citep{sej+13}.  

Though the rate-amplitude constraints drawn in Figure 9 are pessimistic for the chances of an imminent detection of a BWM in light of our estimates for the event rate at a detection threshold of $h=10^{-15}$, PTA sensitivity to BWMs improves strongly with time, scaling approximately as $T^{-3/2}$ until the adverse influences of red timing noise intervene.  Increasing timing baselines by just a factor of two can in principle yield an average BWM sensitivity enhancement by nearly a factor of three.  Also, obviously, the longer time baselines for timing experiments get, the greater the chances of detecting a rare event. Furthermore if $N_P$ pulsars can be timed with a comparable residual rms $\sigma$, the sensitivity of the PTA to BWMs in the Earth term can be enhanced by a factor proportional to $\sigma N_P^{-1/2}$.  The search for BWMs can thus be aided by new and improved timing instruments and techniques that can reduce $\sigma$ and ongoing and future searches for MSPs that can increase $N_P$.  With the continued and combined efforts of the world's PTAs, BWMs with amplitudes at or below $10^{-15}$ will likely become detectable within 5 to 10 years.  Whether or not a detection becomes possible at that point rests on the poorly constrained event rate at that detection threshold.  With an amplitude sensitivity of $10^{-15}$, an event may not be seen for ten thousand years or one may be seen every other year.  Either scenario will teach us something important about the population of SMBHBs.

Finally, we emphasize that memory is a generic feature of any system that emits GWs and that there may be exotic physical processes creating BWMs that have not been incorporated into our rate estimates.  For example, cosmic superstrings, if they exist, almost certainly produce strong GWs in a highly beamed ``rocket"  \citep{c09}, making them potential sources of strong memory.  \citet{cbv+14} have recently identified PTA searches for memory as a means by which GW bursts from early-universe phenomena at extremely high redshifts might be detected. The possibility of detecting such unexpected objects as cosmic strings, learning about the population and evolutionary history of SMBHBs, and potentially detecting the otherwise undetectable signature of some of the most energetic phenomena in the universe through searches for memory is an exciting prospect that is exclusively within the purview of PTAs.\\

We would like to thank the members of the NANOGrav collaboration, especially those in the Detection Working Group.  We additionally thank G. Hobbs, W. Coles, R. Shannon, and Y. Levin, with whom we have had many useful conversations regarding BWMs.  We thank our anonymous referee for a careful reading and helpful comments.  This work was supported by a sub-award to Cornell University from West Virginia University from NSF/PIRE Grant 0968296.

\bibliography{mad0422.bib}
\end{document}